\documentclass[a4paper,fleqn]{cas-dc}
\usepackage{cas-common}
\usepackage{tabu, booktabs}
\usepackage{graphicx}
\usepackage{comment}
\usepackage{float}
\usepackage{datetime}
\usepackage{amsmath,amsfonts,amssymb,mathrsfs}
\usepackage{natbib}
\usepackage{epsfig}
\usepackage{color}
\usepackage{bm}
\usepackage{booktabs}
\usepackage{array}
\usepackage{multirow}
\usepackage{placeins}
\usepackage{caption}
\usepackage{subcaption}
\usepackage{siunitx}
\usepackage{enumitem}
\usepackage{hyperref}
\usepackage{afterpage}
\usepackage{etoolbox}
\usepackage{xcolor}
\usepackage{booktabs} % Para tablas profesionales
\usepackage{amsmath}  % Para ecuaciones
\usepackage[export]{adjustbox} % Para usar valign=t en includegraphics
\usepackage{tabularx}

\DeclareUnicodeCharacter{202F}{FIX ME!!!!}

\def\tsc#1{\csdef{#1}{\textsc{\lowercase{#1}}\xspace}}
\tsc{WGM}
\tsc{QE}
\begin{document}

\let\WriteBookmarks\relax
\def\floatpagepagefraction{1}
\def\textpagefraction{.001}

\title[mode = title]{Multiphysics Finite Element Modeling of Irradiation and Thermal Behavior Demonstrated on a Fuel-Assembly Problem}

%% Short title
\shorttitle{Multiphysics Finite Element Modeling of Irradiation and Thermal Behavior}    
\author[1]{Fabrizio Aguzzi}[]
\author[1]{Mart\'in Armoa}[]
\author[1]{Santiago M. Rabazzi}[]

%\fnmark[1]

\author[1,3]{César Pairetti}[]
\author[2]{Alejandro E. Albanesi}[]
\cormark[1]
\ead{aalbanesi@cimec.santafe-conicet.gov.ar}
\cortext[1]{Corresponding author}

\affiliation[1]{organization={IFIR Instituto de F\'isica de Rosario, CONICET-UNR},
	addressline={Ocampo y Esmeralda, Predio CONICET Rosario}, 
	%          citysep={}, % Uncomment if no comma needed between city and postcode
	postcode={2000}, 
	country={Argentina}}

\affiliation[2]{organization={CIMEC Centro de Investigaci\'on de M\'etodos Computacionales, CONICET-UNL},
	addressline={Col. Ruta 168 s/n, Predio CONICET Santa Fe}, 
	%          citysep={}, % Uncomment if no comma needed between city and postcode
	postcode={3000},
	country={Argentina}}

\affiliation[3]{organization={Sorbonne Université and CNRS},
	addressline={UMR 7190, Institut Jean Le Rond d’Alembert}, 
	city={Paris},
	%          citysep={}, % Uncomment if no comma needed between city and postcode
	postcode={75005}, 
	country={France}}

\begin{abstract}
This work presents a modeling framework to represent the thermomechanical behavior of complex materials, based on micro mechanical dynamics. This tool is then applied to fuel rod elements, consisting of Zircaloy-2 cladding tubes and spacer grids, under typical Pressurized Water Reactor (PWR) conditions. The model incorporates thermal expansion and thermal creep through a VPSC-FEM coupling with the finite element (FEM) solver Code\_Aster, enabling analysis of in-reactor behavior under combined thermal, mechanical, and irradiation loading. The framework captures anisotropic deformation driven by crystallographic texture and prismatic slip activity under radial loading. Thermal creep, being stress-sensitive, contributes to early-stage stress relaxation and strain accumulation, leading to higher strain compared to the irradiation-only case. The interaction of thermal creep with irradiation mechanisms was found to modify the stress distribution and clearance evolution, with relaxation governed by prismatic slip. For fuel rod elements, irradiation-induced mechanisms dominate the long-term clearance behavior, whereas thermal effects remain relevant in contact dynamics during thermal preloading. Furthermore, the stress–strain response was found to be more sensitive to micromechanics than to elasticity. This high-resolution formulation enables predictive modeling of spacer-cladding interaction and provides a basis for the development of reduced-order models.
\end{abstract}

% Research highlights
%\begin{highlights}
%\item Highlight number 1
%\item Highlight number 2
%\item Highlight number 3
%\end{highlights}

% Keywords
% Each keyword is seperated by \sep
\begin{keywords}
Thermal creep \sep Spacer–cladding interaction \sep Zircaloy-2 \sep Polycrystalline modeling \sep Code\_Aster \sep VPSC \sep Thermo-mechanical simulation
\end{keywords}

\maketitle

\section{Introduction} \label{sec:intro}

Understanding the mechanical behavior of zirconium alloys under irradiation is essential for predicting the performance and integrity of nuclear fuel cladding. Deformation in zirconium alloys during in-reactor operation is driven by three concurrent mechanisms: thermal creep, irradiation creep, and irradiation-induced growth \citep{adamson2019irradiation,carpenter1988irradiation,fidleris1988irradiation,holt2008reactor,onimus2020radiation,rogerson1988irradiation}. While irradiation-induced effects are dominant in most high-flux regions, thermal phenomena cannot be neglected, particularly under steady-state or transient conditions where elevated temperatures may activate additional creep mechanisms.

Thermal creep in zirconium alloys typically becomes significant above \( 523\text{--}623 \, \mathrm{K} \), depending on the alloy composition, stress level, and microstructural condition. In unirradiated materials, this phenomenon is governed by well-known diffusional processes. However, under neutron irradiation, the microstructural evolution substantially modifies the creep response. This process involves the formation of point defects, loops, and dislocation structures. As a result, extrapolations from ex-reactor data may be unreliable. Moreover, in low-flux regions (e.g., near fuel rod extremities), thermal creep can remain active for extended periods, contributing significantly to strain accumulation \citep{adamson2009reactor, adamson2019irradiation}.

Moore et al. \citep{moore2025stress} emphasized the intricate interplay between irradiation and thermal creep, noting that these mechanisms—although often assumed additive and independent—operate simultaneously under reactor conditions, complicating the interpretation of dimensional changes. While irradiation creep is generally considered athermal, weakly dependent on temperature, thermal creep becomes increasingly relevant at elevated temperatures. This distinction is particularly important in scenarios involving Pellet-Cladding Mechanical Interaction (PCMI), such as during power ramps, where localized overheating can trigger thermal creep. These observations highlight the pressing need for comprehensive, physically-based models that incorporate both irradiation and thermal effects to accurately capture their respective contributions to fuel cladding deformation.

Despite extensive work on creep modeling in zirconium alloys, many contributions have focused on pressure tubes in heavy-water reactors or on cladding tubes under simplified loading conditions. Phenomenological macroscopic models, derived from experimental campaigns, have been widely used to study internal/external pressure, temperature, anisotropy, and irradiation effects \citep{murty1999creep,matsuo1987thermal}. However, these models are often limited in scope and typically valid only within specific parameter ranges. Several researchers \citep{christodoulou1996modeling,limback1996model} have proposed empirical approaches where deformation mechanisms are added at the macroscopic level without explicitly accounting for the polycrystalline structure of the material.

Due to their technological importance and strong plastic anisotropy, zirconium alloys have received considerable attention since the seminal work by \cite{woo1985polycrystalline}. Subsequent studies explored the anisotropic response under irradiation using polycrystalline models and homogenization techniques, focusing on irradiation creep and growth \citep{carpenter1988irradiation,lebensohn1993self,turner1993self,tome1993self,turner1994self,tome1996role,turner1999self,christodoulou2002analysis,patra2017finite,montgomery2017use}. Although similar micromechanical approaches have also been proposed for modeling thermal creep \citep{brenner2002quasi}, their application remains limited.

More recently, Fast Fourier Transform (FFT)-based methods have been employed to study the combined effects of irradiation and thermal mechanisms in polycrystalline aggregates of Zircaloy-4 \citep{onimus2020radiation,onimus2022polycrystalline,gicquel2023polycrystalline}. These grain-level simulations reinforce the need for physically-based models capable of capturing microstructural effects, crystallographic texture, and the directional nature of slip and growth processes.

Building upon this perspective, the present work introduces an extended polycrystalline constitutive framework based on the VPSC model, coupled with the finite element code Code\_Aster \citep{ASTER} via the CAFEM-VPSC interface \citep{aguzzi2025toolbox}. The current model incorporates two additional thermally activated mechanisms, thermal creep and thermal expansion, enabling a more comprehensive and realistic simulation of in-reactor deformation in Zircaloy-2.

In section 3 of this manuscript, this framework is employed to simulate the non-linear evolution of the clearance between the cladding tube and spacer grid under Pressurized Water Reactor (PWR) conditions. This analysis aims to assess the relative importance of thermal and irradiation effects by applying a comprehensive material description based on a micromechanical approach. In section 4, a detailed description of the stress and strains shows the impact of each micromechanical phenomenon on the global behavior of the assembly. By capturing the coupled response of Spacer-Cladding Interaction (SCI), the study enhances the predictive capabilities of computational tools for evaluating fuel rod performance and structural integrity under realistic reactor operation.

\FloatBarrier
\section{Multiphysics Simulation Methodology}\label{sec:case}

\subsection{Micro-Scale Model}
\label{EcCristalSimple}
\subsubsection{Irradiation-Induced Growth}
We apply a model based in a reaction-diffusion formulation, following \cite{patra2017crystal}.
In this context, the strain-rate tensor due to irradiation-induced growth is obtained by linear superposition over the crystallographic directions $j \in \{\mathbf{a}_1,\mathbf{a}_2,\mathbf{a}_3,\mathbf{c}\}$:
\begin{equation}\label{EcuacionCrecimiento}
	\dot{\varepsilon}_{kl}^{(\text{growth})} \;=\;
	\sum_{j \,\in\,\{\mathbf{a}_1,\mathbf{a}_2,\mathbf{a}_3,\mathbf{c}\}}
	\dot{\varepsilon}_{\text{growth}}^j \;b_k^j\,b_l^j,
	\quad
	k,l \,\in\,\{\mathbf{x},\mathbf{y},\mathbf{z}\},
\end{equation}
where $b_k^j$ and $b_l^j$ are the components of the Burgers vector $\mathbf b^j$ along the crystal‐axis directions $\mathbf k$ and $\mathbf l$, respectively. Each deformation rate $\dot{\varepsilon}_{\text{growth}}^j$ is the sum of effects from dislocation‐climb and grain‐boundary‐absorption induced growths:
\begin{equation}
	\dot{\varepsilon}_{\text{growth}}^j \;=\;
	\dot{\varepsilon}_{\text{climb}}^j \;+\; \dot{\varepsilon}_{\text{GB}}^j,
	\quad
	j \,\in\,\{\mathbf{a}_1,\mathbf{a}_2,\mathbf{a}_3,\mathbf{c}\},
\end{equation}
where the $\dot{\varepsilon}_{\text{climb}}^j$ is defined based on the defect production rate from vacancies and interstitials, following ~\cite{norgett1975proposed}, and $\dot{\varepsilon}_{\text{GB}}^j$ follows an equivalent rationale for grain boundaries, as described in \cite{aguzzi2025toolbox}.

\subsubsection{Irradiation-induced creep crystallography
	model}

The growth of one grain under irradiation can be interrupted by the growth of a neighbouring grain. This incompatibility creates internal stresses known as creep under irradiation. Irradiation creep is intrinsically coupled with irradiation-induced growth, occurring independent of external loads.

Our analysis uses the crystallographic irradiation creep model for steels (~\cite{foster1972analysis},~\cite{ehrlich1981irradiation}) as implemented by \cite{patra2017crystal}, with swelling effects disregarded. This constitutive law models the creep strain rate as linearly proportional to both applied stress and radiation dose rate \( \frac{d\phi}{dt} \), expressed in terms of effective stress and strain components:

\begin{equation}
	\label{TasaCreep}
	\dot{\varepsilon}^{\text{creep}} = B_{0} \sigma \frac{d\phi}{dt};
\end{equation}

\noindent effective stress is denoted by \( \sigma \), \( B_0 \) is the irradiation creep compliance, and \( \frac{d\phi}{dt} \) is the radiation dose rate (in \( dpa\cdot s^{-1} \)). A similar phenomenological law is assumed to represent the shear rate associated with irradiation creep at the level of slip systems, as:

\begin{equation}
	\label{TasaCorteCreep}
	\dot{\gamma}_{\text{creep}}^{\text{j}} = B \frac{\rho_d^j}{\rho_\text{ref}} \tau^j \frac{d\phi}{dt};
\end{equation}

\noindent where \( B \) is the crystallographic irradiation creep compliance, \( \tau^j \) is the resolved shear stress on slip system \( j \), and \( \rho_\text{ref} \) is the reference line dislocation density. The crystallographic shear rate is a function of the resolved shear stress \( \tau^j \) and of the line dislocation density \( \rho_d^j \) gliding on slip system \( j \). The anisotropic creep arises from the distribution of dislocations among the slip systems, such that the creep strain rate (and its associated relaxation) is greater in systems with higher dislocation density. The normalization factor \( \rho_\text{ref} \) ensures that the crystallographic irradiation creep compliance \( B \) has the same physical units as the macroscopic compliance \( B_0 \) used in Eq.~\ref{TasaCreep}. The specific values for the irradiation creep model parameters adopted in this work are provided in Appendix~\ref{apéndiceA} and were taken from \cite{patra2017crystal}.

\subsubsection{Thermal Creep Model}

Thermal creep is a phenomenon associated with the motion of dislocations that overcome obstacles via temperature- and stress-activated diffusion mechanisms. Although this process becomes more pronounced at elevated temperatures (\(T > 0.5 T_M\), being $T_M$ the melting temperature), or under stresses close to the yield strength (\(\sigma > 0.75\sigma_f\)), it is not limited to abnormal or accidental conditions. Experimental tests under steady-state conditions have demonstrated that this mechanism can also be activated within the normal operating range of nuclear reactors, with significant effects on material behavior~\citep{christodoulou2002analysis,fidleris1988irradiation,xiao2023modelling}.

The power-law expression in Eq.~\ref{EcuacionThermalCreep} is suitable for describing the contribution of multiple slip systems to the total creep within a grain:
\begin{equation}
	\label{EcuacionThermalCreep}
	\dot{\varepsilon }_{ij}^{\text{thcreep}} = \sum_s m_{ij}^s \dot{\gamma^s} = \dot{\gamma_0} (T) \sum_s m_{ij}^s
	\left(\frac{m_{ij}^s \sigma_{\text{kl}}}{\tau_{\text{thres}}^s}\right)^n;
\end{equation} 

In the expression given by Eq.~\ref{EcuacionThermalCreep}, \(m_{ij}^s\) is the Schmid tensor, defined as \mbox{\(m_{ij}^s = \frac{1}{2} (b_i^s n_j^s + b_j^s n_i^s)\)}, where \(\boldsymbol{n}^s\) is the normal to the slip plane, and \(\boldsymbol{b}^s\) is the slip direction (Burgers vector) of the \(s\)-th slip system. Slip may occur on prismatic, basal, or pyramidal planes (see Section~\ref{VPSC}). The parameters in Eq.~\ref{EcuacionThermalCreep} are related to the physical properties of Zr alloys and are detailed in the following paragraphs.

The reference shear rate \(\dot{\gamma_0}\) depends on temperature and the activation energy of the diffusion mechanism that controls the process. For this study, we adopt the temperature dependence proposed in \cite{christodoulou2002analysis}, derived from the fitting of an extensive experimental database from tensile and shear creep tests on Zr-2.5\%Nb alloy tubes under various stresses, temperatures, and directions. In that work, \(\dot{\gamma_0}\) is assumed identical for all slip systems, an idealization, and the creep behavior of each system is distinguished by its threshold stress \(\tau_\text{thres}^s\). Those authors report:

\begin{equation}
	\label{gammadotT}
	\dot{\gamma}_0(T) = \dot{\gamma}_0(T=523K) \exp\left[-Q(T)\left(\frac{1}{T}-\frac{1}{523}\right)\right],
\end{equation} 

\noindent where \(\dot{\gamma_0}(T=523K)=4.157 \times 10^{-7} \ h^{-1}\), and 

\begin{equation}
	\label{QT}
	Q(T) = \left( 5000 + \frac{5600}{1 + \exp^{-(\frac{T-470}{15})}} \right),
\end{equation} 

Experimental studies report a wide range of thermal creep parameters for zirconium alloys, reflecting their sensitivity to alloy composition, temperature, and stress levels. The rate sensitivity exponent \(n\) typically ranges from 3.4 to 5.0, although values as high as 100 have been reported under high-stress regimes~\citep{christodoulou2002analysis,wang2013mechanism,hayes2006creep,franklin2003}. Pre-exponential factors \(\dot{\gamma}_0\) also vary significantly, with values on the order of \(10^{-8} \, \text{s}^{-1}\) reported in the literature~\citep{onimus2022polycrystalline}. These variations highlight the need to calibrate model parameters according to the specific material and loading conditions.

In this work, we adopt \(n = 4\) and use threshold resolved shear stresses of 0.100~GPa (prismatic), 0.111~GPa (basal), and 0.300~GPa (pyramidal). Other experimental datasets, such as those in~\citep{butcher1986}, can be used to fit and validate the model under irradiation conditions.

\subsubsection{Crystallographic Model for Thermal Expansion}  
\label{model due to Thermal Expansion}

A temperature change induces a resulting strain in the material. Under certain conditions, the components of the thermal strain tensor are proportional to the temperature increment:

\begin{equation}
	\label{Ec.ExpansiónTérmica}
	\varepsilon^\text{therm}_{ij} = \alpha_{ij}\Delta T    
\end{equation}

\noindent where \(\alpha_{ij}\) are the thermal expansion coefficients of the material. These coefficients are generally positive, although negative values have been reported in specific cases~\citep{sleight1998isotropic}. Moreover, thermal expansion coefficients are strongly temperature dependent.

The thermal expansion coefficients used for zirconium in this study were \(\alpha_{\text{A;Ro}} = 5.4\), \(\alpha_{\text{H;Tr}} = 9.1\), and \(\alpha_{\text{R;No}} = 7.4\), all expressed in units of \(10^{-6} \ \text{K}^{-1}\). The subscripts refer to the axial (A), hoop (H), and radial (R) directions of the cladding tube, as well as the rolling (Ro), transverse (Tr), and normal (No) directions of the spacer grid dimples, which are defined according to the crystallographic texture shown in Fig.~\ref{fig:ReducedTexture}. These values are based on experimental measurements reported by \cite{goldak1966lattice}.

For the simulations, the elastic constants reported by \cite{simmons1965single,kocks2000texture} were adopted. At room temperature, the values used were:

\begin{equation*}
	\begin{aligned}
		C_{11} &= 143.5 \ \text{GPa}, \quad 
		C_{33} = 164.9 \ \text{GPa}, \\
		C_{44} &= 32.1 \ \text{GPa}, \quad 
		C_{12} = 72.5 \ \text{GPa}, \quad 
		C_{13} = 65.4 \ \text{GPa}.
	\end{aligned}
\end{equation*}

These values reflect the strong elastic and thermal anisotropy of zirconium, which plays a key role in transient scenarios and must be properly accounted for in predictive modeling.

\subsection{Meso-Scale Model} 
\subsubsection{Viscoplastic Self-Consistent Model (VPSC)}
\label{VPSC}

The VPSC model \citep{molinari1987self, tome1993self} has been extended over time to include creep and irradiation growth at the grain scale, considering texture and grain interactions \citep{patra2017crystal, tome2023material}. In this work, it is further developed to incorporate thermal creep and and thermal expansion, given by Eqs. \eqref{EcuacionThermalCreep}, \eqref{Ec.ExpansiónTérmica}, respectively, enhancing its ability to simulate the thermomechanical behavior of zirconium alloys under reactor conditions.

The total deformation rate at the grain level is given by the contribution of the phenomena mentioned above: (\ref{EcuacionCrecimiento},\ref{TasaCreep},\ref{EcuacionThermalCreep},\ref{Ec.ExpansiónTérmica}):

\begin{equation}
\dot{\varepsilon}_{ij}=\dot{\varepsilon}_{ij}^\text{(growth)}+\dot{\varepsilon}_{ij}^\text{(creep)}+\dot{\varepsilon}_{ij}^\text{(thcreep)}+\dot{\varepsilon}_{ij}^\text{(therm)}
\end{equation}

The VPSC code can be executed under different linearization schemes, and the corresponding results may differ significantly when using the same hardening parameters~\cite{tome2023material}. For the response of the effective medium (polycrystal), the tangent linearization was used in this work:

\begin{equation} \bar{\dot{\varepsilon}}_{ij}=\bar{M}_{ijkl} \bar{\sigma}_{kl}+\bar{\dot{\varepsilon}}_{ij}^0
\end{equation}

\noindent where \(\bar{\dot{\varepsilon}}_{ij}\) and \(\bar{\dot{\sigma}}_{kl}\) represent the average (macroscopic) strain rate and stress, respectively. \(\bar{M}_{ijkl}\) denotes the macroscopic creep compliance, while \(\bar{\dot{\varepsilon}}_{ij}^0\) corresponds to the macroscopic strain rate resulting from irradiation-induced growth. These macroscopic quantities, \(\bar{M}_{ijkl}\) and \(\bar{\dot{\varepsilon}}_{ij}^0\) are initially unknown and must be determined self-consistently by enforcing that the volume-averaged stress and strain rate over all grains match those of the effective medium, namely:

\begin{equation}
    \bar{\dot{\varepsilon}}_{ij}=\langle \dot{\varepsilon}_{ij}^c \rangle; \ \ \bar{\sigma}_{kl}=\langle \sigma_{ij}^c \rangle
\end{equation}

Previous applications of the self-consistent polycrystalline model focused on simulating creep and irradiation-induced growth in hexagonal materials \cite{tome1996role, turner1999self}, but did not incorporate the time evolution of microstructural changes. In this study, we build upon that framework by introducing time-dependent formulations for both creep and irradiation growth. While \cite{patra2017finite} implemented this on a proprietary FEM solver, our approach is implemented within Code\_Aster \cite{ASTER}, an open-source FEM solver suited for nuclear engineering problems. 

Regarding thermal phenomena, \cite{montgomery2017use} incorporated thermal creep but did not consider thermal expansion. In contrast, our current formulation includes both thermal creep and thermal expansion at the grain level. In summary, the toolbox integrating Code\_Aster and VPSC to model irradiation creep and growth, developed in \cite{aguzzi2025toolbox}, has been expanded to also incorporate thermal creep and thermal expansion, resulting in a more versatile and comprehensive model to analyze the nonlinear contact between grid and cladding.

\subsubsection{Sensitivity Analysis of Thermal Effects}
\label{sec: Deformation mechanisms influencing gap opening}

Although thermal creep and thermal expansion typically induce smaller strains than irradiation effects, their contribution becomes relevant under reactor operating conditions due to elevated temperatures and their role in stress relaxation. These thermally activated mechanisms help redistribute stresses and influence the overall deformation behavior, particularly in textured materials such as Zircaloy-2.

\begin{figure}[h!] 
    \centering
    \includegraphics[width=0.45\textwidth]{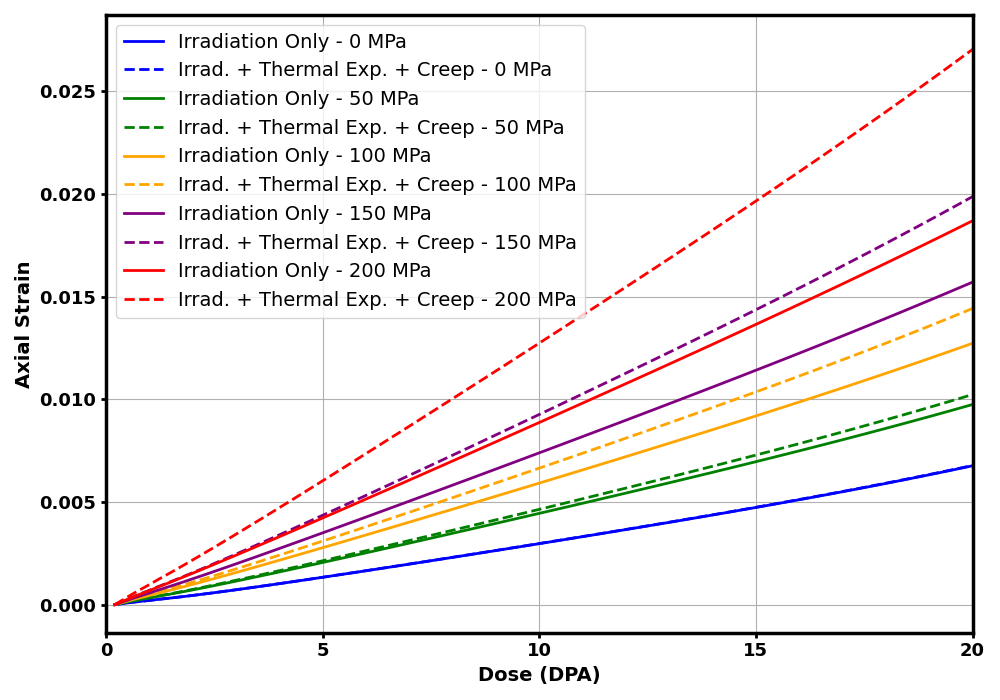}
    \caption{Axial strain as a function of axial load for the texture shown in Fig. \ref{fig:ReducedTexture}(a), considering either irradiation creep alone or the combined effects of irradiation creep and thermal creep (including thermal expansion for heating rate\( =0.0018K/h\)). }
    \label{fig: Axial Strain}
\end{figure}

Figure~\ref{fig: Axial Strain} highlights the dependence of thermal creep on the applied stress to the cladding tube. The curves show the axial strain evolution at 523~K for different axial loads (0, 50, 100, 150, 200~MPa). Solid lines represent the strain induced by irradiation effects only (creep and growth), while dashed lines show the total axial strain when thermal creep and thermal expansion are also included. These results were obtained using the VPSC stand-alone (VPSC-SA) code, without integration to the finite element solver. The results clearly demonstrate that, for a given stress level, the total strain is consistently higher when thermal effects are considered—indicating the stress-sensitive nature of thermal creep.

This observation is consistent with the thermal creep formulation in Eq.~\ref{EcuacionThermalCreep}, where the strain rate is proportional to the stress raised to a power \(n\) (typically \(n=4\) for Zircaloy). Although the temperature dependence enters through the prefactor \(\dot{\gamma}_0(T)\) in Eq.~\ref{gammadotT}, its exponential variation is much slower compared to the polynomial amplification of stress. This illustrates that, under typical PWR conditions, thermal creep is significantly more sensitive to stress than to temperature variations.

This behavior is closely related to the crystallographic texture. As shown in Fig.~\ref{fig:ReducedTexture}(a), the predominance of prismatic planes aligned with the axial and hoop directions facilitates thermal creep along those directions due to their lower critical resolved shear stress. Under compressive radial loading (e.g., from external pressure), these slip systems efficiently accommodate plastic deformation, thereby contributing to stress relaxation.

\subsection{Macro-Scale Model} 

\subsubsection{VPSC-CAFEM interaction framework}
\label{Acople VPSC-FE}

The VPSC-CAFEM interface integrates the polycrystalline constitutive model VPSC with the finite element solver Code\_Aster \citep{aguzzi2025toolbox}. At each time increment, the solver provides the strain increment \(\Delta \boldsymbol{\varepsilon}_{\text{FE}} \), the time step \(\Delta t\), and a rotation matrix \( R \) defining the local coordinate system.

The total strain increment is decomposed into elastic and viscoplastic contributions:
\begin{equation}
\Delta\boldsymbol{\varepsilon} = \Delta \boldsymbol{\varepsilon}^{\text{e}} + \Delta \boldsymbol{\varepsilon}^{\text{vp}} = \mathbf{C}^{-1}:\Delta\boldsymbol{\sigma} + \Delta \boldsymbol{\varepsilon}^{\text{vp}}
\end{equation}

The viscoplastic strain results from mechanisms such as irradiation creep and thermal creep (stress-dependent), as well as irradiation growth and thermal expansion (stress-independent). These contributions depend on the current stress \(\boldsymbol{\sigma}\) and internal state variables, such as grain orientations and hardening parameters.

To ensure consistency with the local frame used in Code\_Aster, stress and strain tensors are rotated using:
\begin{equation}
\Delta\boldsymbol{\varepsilon}^* = \boldsymbol{R} \Delta\boldsymbol{\varepsilon} \boldsymbol{R}^T, \quad \boldsymbol{\sigma}^* = \boldsymbol{R} \boldsymbol{\sigma} \boldsymbol{R}^T
\end{equation}

The superscript \(^*\) indicates quantities in the local coordinate system. A Newton-Raphson iterative scheme is employed to solve for the updated stress. The residual function, defined as the difference between the strain of the constitutive model and the strain provided by the FEM solver \citep{segurado2012multiscale}, is given by:

\begin{equation}
    \mathbf{X}\left( \Delta\boldsymbol{\sigma}^*   \right) = \Delta\boldsymbol{\varepsilon}^*-\Delta\boldsymbol{\varepsilon}^*_{FE}=\mathbf{C}^{-1}:\Delta\boldsymbol{\sigma}^*+\boldsymbol{\dot{\varepsilon}}^{vp*}\Delta t- \Delta\boldsymbol{\varepsilon}^*_\text{FE}
\end{equation}

The stress increment is updated iteratively as:
\begin{equation}
\label{ec:correcciondeTension}
    \left( \Delta\boldsymbol{\sigma}^*\right)_{k+1}=\left( \Delta\boldsymbol{\sigma}^*\right)_{k}-\textbf{J}^{-1^*}_{NR}((\Delta\boldsymbol{\sigma}^*)_{k}):\textbf{X}((\Delta\boldsymbol{\sigma}^*)_{k})
\end{equation}

\noindent where the Newton-Raphson Jacobian 
\(\textbf{J}^*_{NR}=\textbf{C}^{-1}+\textbf{M}\Delta t\) includes the elastic stiffness \(\textbf{C}\) and the viscoplastic tangent modulus \(\textbf{M}\) provided by the VPSC model.

Upon convergence, the stress tensor and the consistent tangent operator are rotated back to the global coordinate system to ensure compatibility with the Code\_Aster solver:

\begin{equation}
\boldsymbol{\sigma}^{t+\Delta t} = \boldsymbol{R}^T \boldsymbol{\sigma}^{t+\Delta t^*} \boldsymbol{R}, \quad 
\mathbf{C}^{\text{tg}} = \mathcal{R} \, \mathbf{C}^{\text{tg*}} \, \mathcal{R}^T
\end{equation}

Here, \(\mathcal{R}\) denotes the fourth-order rotation tensor derived from the second-order rotation matrix \(\boldsymbol{R}\). The component-wise transformation of the consistent tangent operator is given by:

\begin{equation}
C^{\text{tg}}_{ijkl} = R_{im} R_{jn} R_{ko} R_{lp} \, C^{\text{tg*}}_{mnop}
\end{equation}

Convergence is monitored by evaluating a weighted norm of the residual \citep{mcginty2001multiscale}:
\begin{equation}
\chi = \sqrt{\sum_{i,j} \left( \frac{|\Delta\varepsilon_{\text{FE}}^{ij}|}{\max(|\Delta\varepsilon_{\text{FE}}^{ij}|)} X^{ij} \right)^2}
\end{equation}

A validation of this coupling strategy, including comparison with stand-alone VPSC results, is provided in Appendix~\ref{apéndiceB}.

\section{Mechanical Contact Analysis in Spacer Grid-Cladding Systems}  
\label{sec:simulation-setup}

\begin{figure}[h] 
    \centering
    \includegraphics[width=0.5\textwidth]{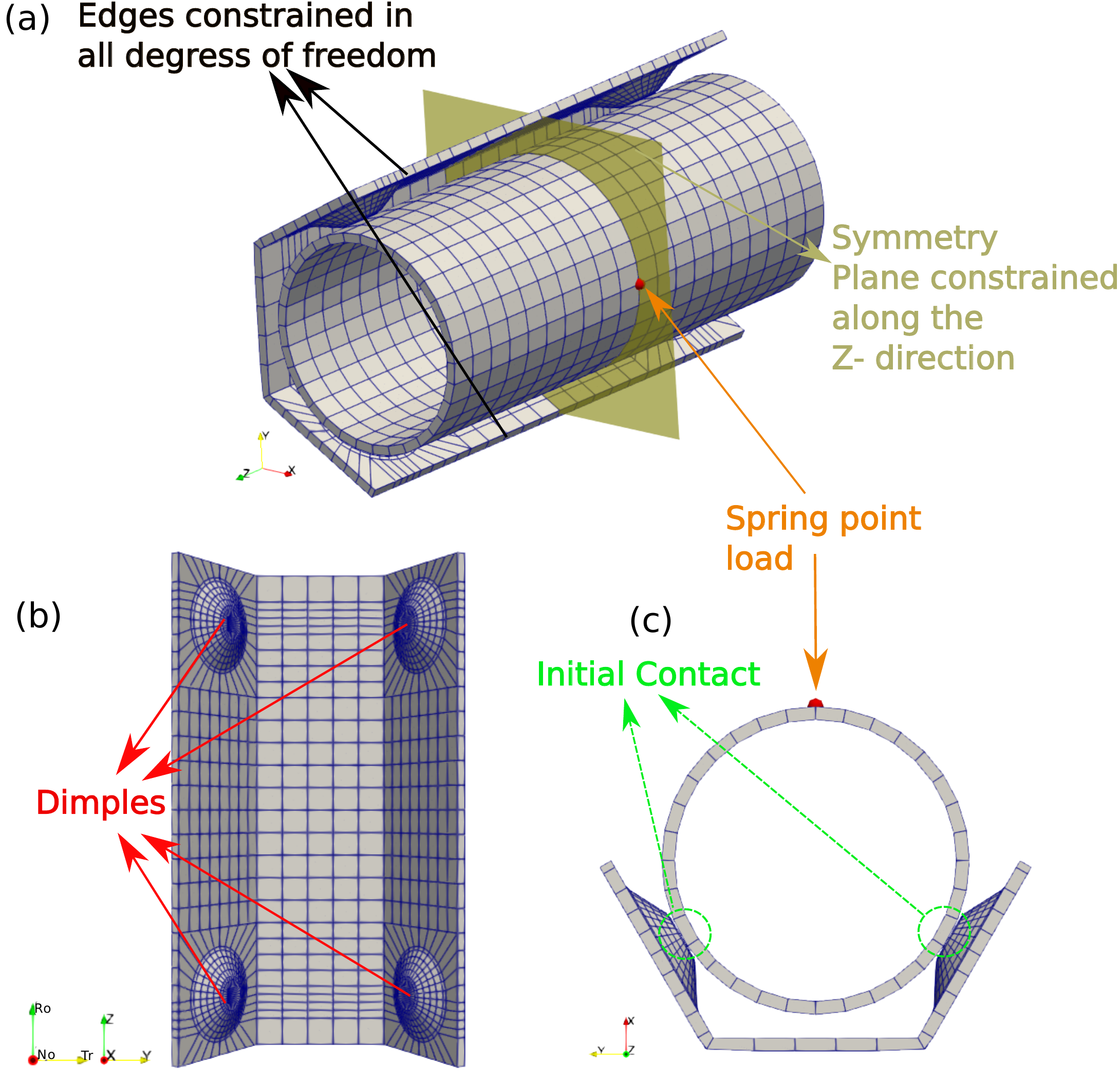}
    \caption{(a) Cladding tube–grid assembly. (b) Front view of the grid with four supporting dimples. The bottom-left corner shows the coordinate system for the grid sheet. (c) Top view showing the spring contact force on the cladding tube.}
    \label{fig: MallaTuboyChapa}
\end{figure}

The simulation domain consists of a reduced model comprising half of the cladding tube and a portion of the central hexagon of the spacer grid, a common design in PWRs. Representative images of the spacer grid structure are available on the official CONUAR website~\citep{CONUAR}. This simplification, based on geometric symmetry (Fig.~\ref{fig: MallaTuboyChapa}), significantly reduces computational cost while maintaining the physical representativeness of the contact problem. The spacer grid includes four dimples that interact with the tube surface, and a representative spring force of 45~N ensures initial mechanical contact.

The assembly was subjected to constant in-reactor conditions: 523~K temperature, 15.5~MPa external coolant pressure, and 10~MPa internal pressure. A uniform irradiation dose rate of \(3.6 \times 10^{-4} \ \text{dpa} \cdot \text{hr}^{-1}\) was applied for 100 increments of 555.6 hours each. No additional external loads were imposed; however, the pressure differential introduces circumferential and axial stresses in the structure.

\begin{figure}[h] 
    \centering
    \includegraphics[width=0.45\textwidth]{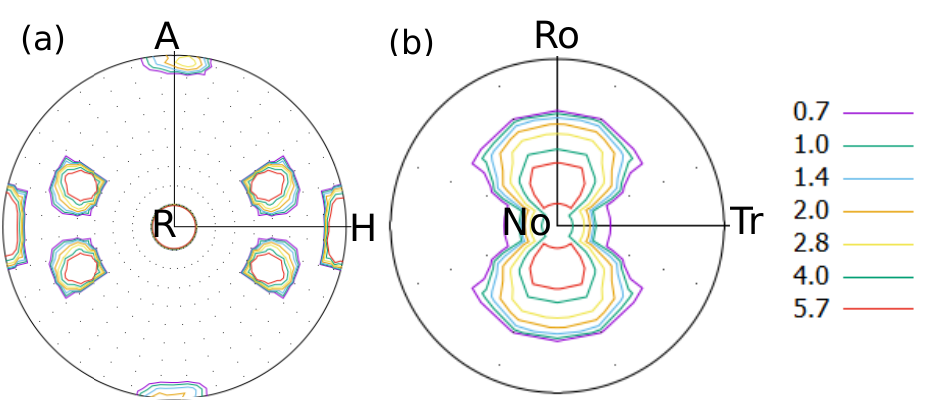}
    \caption{Basal pole figures of the reduced textures used in the simulation. (a) Cladding tube: 7 orientations aligned with axial (A), hoop (H), and radial (R) directions. (b) Spacer dimples: 13 orientations with rolling (Ro), transverse (Tr), and normal (No) directions.}
    \label{fig:ReducedTexture}
\end{figure}

To balance fidelity and efficiency in the VPSC-CAFEM coupling, a reduced texture approach was used~\citep{aguzzi2025toolbox,patra2017finite}, assigning 7 grain orientations to the cladding tube and 13 to the spacer dimples (Fig.~\ref{fig:ReducedTexture}). The cold-worked Zircaloy-2 model was calibrated against irradiation growth data at 550~K~\citep{holt1996non}.

Boundary conditions imposed axial constraints at the base of the tube (allowing radial displacement), and full fixation of the lateral dimple-edge nodes to eliminate rigid body motion. A preliminary elastic step ensured contact formation without irradiation or thermal loads. Once established, the simulation switched to a nonlinear contact model including irradiation and thermal phenomena, using crystallographic textures.

Contact was modeled using a node-to-segment master–slave algorithm implemented in Code\_Aster \citep{CodeAster_U20404}. The cladding tube was defined as the master surface and the grid as the slave. Friction was neglected to isolate mechanical effects. This strategy enables accurate capture of local displacements and stress gradients—improving upon classical methods based on uniform displacement or predefined gap—and allows the clearance (CLR) variable to track component separation throughout the simulation.

\section{Numerical results and analysis} \label{sec:resu}
This section presents the results obtained with the open-source toolbox that integrates Code\_Aster and VPSC to model irradiation creep and growth, and in particular the implementation of thermal creep and thermal expansion.

\subsection{Evolution of contact opening}
\label{Evolution of contact opening}

\begin{figure}[h!] 
	\centering
	\includegraphics[width=0.45\textwidth]{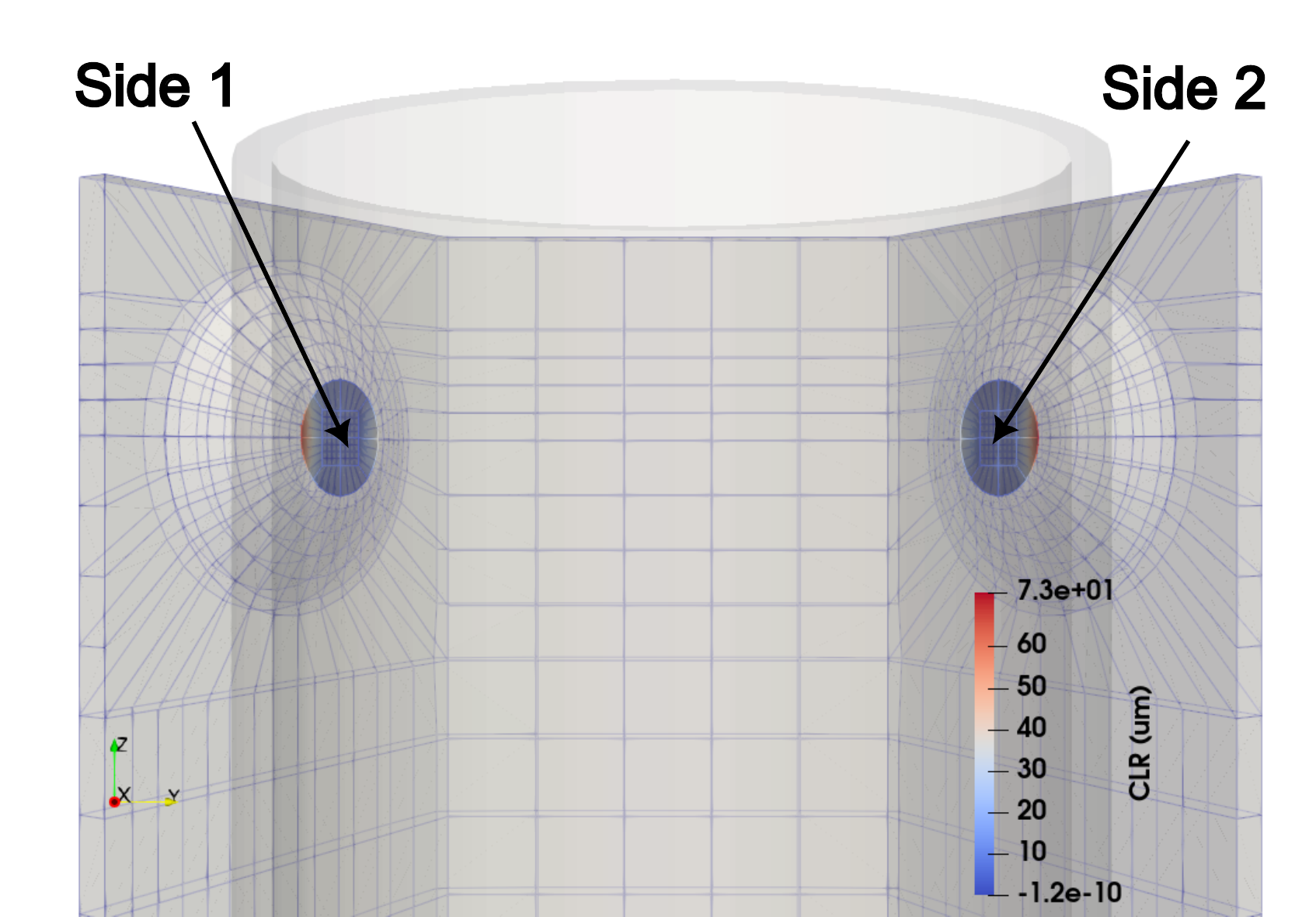}
	\caption{CLR corresponding to the slaves surfaces for non-linear contact for 20 dpa and 523K.
	}
	\label{fig:Side1andSide2}
\end{figure}

In this study, a nonlinear contact analysis is conducted using Code\_Aster to enable a node‑by‑node evaluation of contact zones, accounting for differential displacements and stress distributions at meso$‑$ and macroscopic scales. Such modeling approach significantly improves the accuracy of predictions for contact stresses and separation phenomena compared to conventional rigid-body frameworks, as it captures spatially resolved interactions governed by material and geometric nonlinearities. Fig.~\ref{fig:Side1andSide2} shows the contact surfaces under analysis. Considering an orientation in which the X-axis emerges perpendicularly from the plane, the contact located on the left is referred to as "Side 1", while the contact on the right is identified as "Side 2".

Figure~\ref{fig:acople} shows the evolution of the maximum CLR under different physical mechanisms. The curves represent the individual and combined contributions of irradiation growth, irradiation creep, thermal expansion, and thermal creep. All simulations were performed using the crystallographic textures shown in Fig.~\ref{fig:ReducedTexture}.

This value corresponds to the highest CLR among all nodes, specifically at the peripheral nodes of the slave surface of the dimple, where the separation from the cladding tube is greatest (see red area in Fig. ~\ref{fig:Side1andSide2}). The cladding tube element in contact with the dimple surface undergoes a pivoting motion around the region of highest contact (and therefore highest stress) until it fully adjusts. At a dose above 1.2 dpa, the components of slides 1 and 2 begin to separate due to the deformation mechanisms described in Section ~\ref{sec: Deformation mechanisms influencing gap opening} for the irradiation and thermal creep. %If thermal effects are the only ones considered, the cladding tube and grid begin to move closer to each other because of thermal expansion (see Sec \ref{Influence of the thermal effects on CLR}).

Although experimental data on in-reactor CLR remains scarce, prior reactor investigations typically report values between 10 and 30~µm for grid-to-rod contact~\cite{joulin2002effects, jiang2016grid, billerey2005evolution, patra2017finite}.  These values refer to the change in clearance, i.e., the difference between the final and initial separation. The present work extends previous modeling approaches by integrating thermal effects and capturing the combined contribution of irradiation growth, irradiation creep, thermal expansion, and thermal creep at the crystal scale. By explicitly resolving crystallographic textures and slip system activity, the model offers a more physically realistic prediction of CLR evolution under operational conditions, reinforcing its potential as a robust multiscale tool for spacer-cladding interaction analysis.

\subsection{Combined thermal and irradiation effects}

The analysis of this section studies the influence of thermal and irradiation-induced phenomena on the contact and separation behavior between the cladding tube and the spacer grid. To this end, five numerical simulations were performed: four isolating each physical phenomenon individually, and one combining all phenomena simultaneously.

Although from a physical point of view the decoupling of these mechanisms is not strictly realistic—since all are inherently present and interacting during operation—this separation allows for a clearer understanding of the individual contribution of each mechanism to the evolution of the CLR.

\begin{figure}[h!] 
    \centering
    \includegraphics[width=0.51\textwidth]{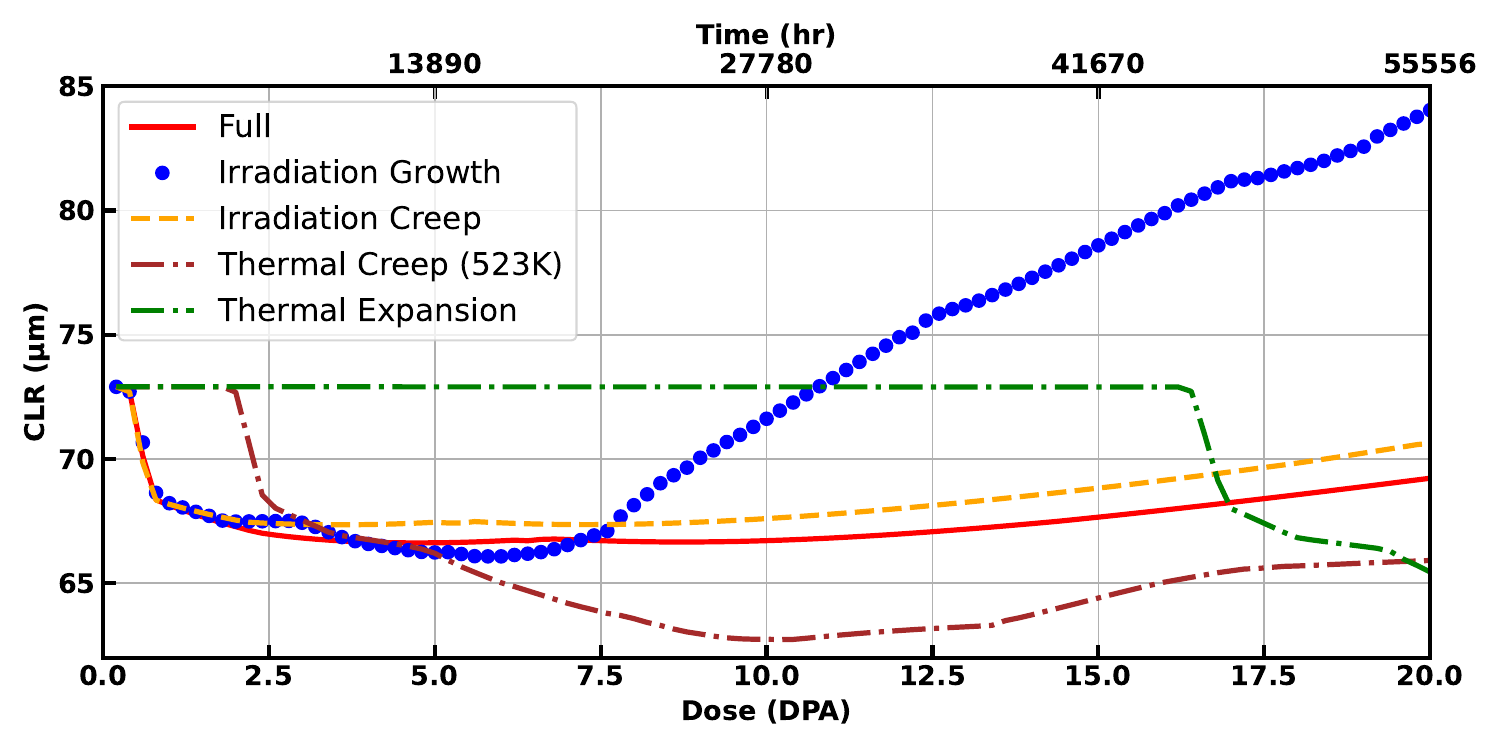}
    \caption{Evolution of CLR distance under different physical mechanisms. The CLR is evaluated by isolating each phenomenon individually: irradiation growth (blue symbols), irradiation creep (orange dashed line), thermal expansion (green dash-dotted line), thermal creep (brown dash-dotted line), and the combined case with all active phenomena — full (red solid line).}
    \label{fig:acople}
\end{figure}

Figure~\ref{fig:acople} summarizes the individual and combined effects of the physical mechanisms on CLR evolution. The simulation with only \textit{irradiation growth} (blue symbols) shows a rapid initial approach between tube and grid, stabilizing up to approximately 7.5~dpa, after which a linear separation trend begins. This is attributed to anisotropic lattice deformation: contraction along the $c$-axis and expansion along the $a$-axes of the HCP zirconium lattice. As shown in Fig.~\ref{fig:ReducedTexture}, the alignment of basal poles enhances this radial expansion.

In the case of \textit{thermal expansion}, a constant heating rate of $1.799 \times 10^{-5}$~K/h (i.e., $\Delta T = 1$~K over the simulation) was applied. The CLR remains nearly constant for about 45,000 hours, followed by a slight reduction (down to ~65~$\mu$m), consistent with thermal softening of the cladding material.

The simulation with \textit{thermal creep} shows a progressive approach between tube and grid during the first 7,000 hours due to positive radial creep strain. This continues until ~27,780 hours when mechanical contact is achieved at the dimple (see Fig.~\ref{fig:CombinedAnalysis}(j-l)). Beyond this, the contact constrains deformation, and a slight separation forms on the opposite side due to bending effects.

In the case of \textit{irradiation creep}, the CLR shows a quasi-linear increase once equilibrium is reached at approximately 1.2~dpa. This linear trend suggests that irradiation creep strain develops gradually and consistently with the applied load.

When \textit{all phenomena are considered simultaneously}, the evolution is dominated by irradiation-induced effects, which produce significantly greater strain than thermal mechanisms. After the initial accommodation (~1.2~dpa), a nearly linear separation trend is observed, reflecting the cumulative contribution of both irradiation growth and creep.

\begin{figure}[h!] 
    \centering
    \includegraphics[width=0.5\textwidth]{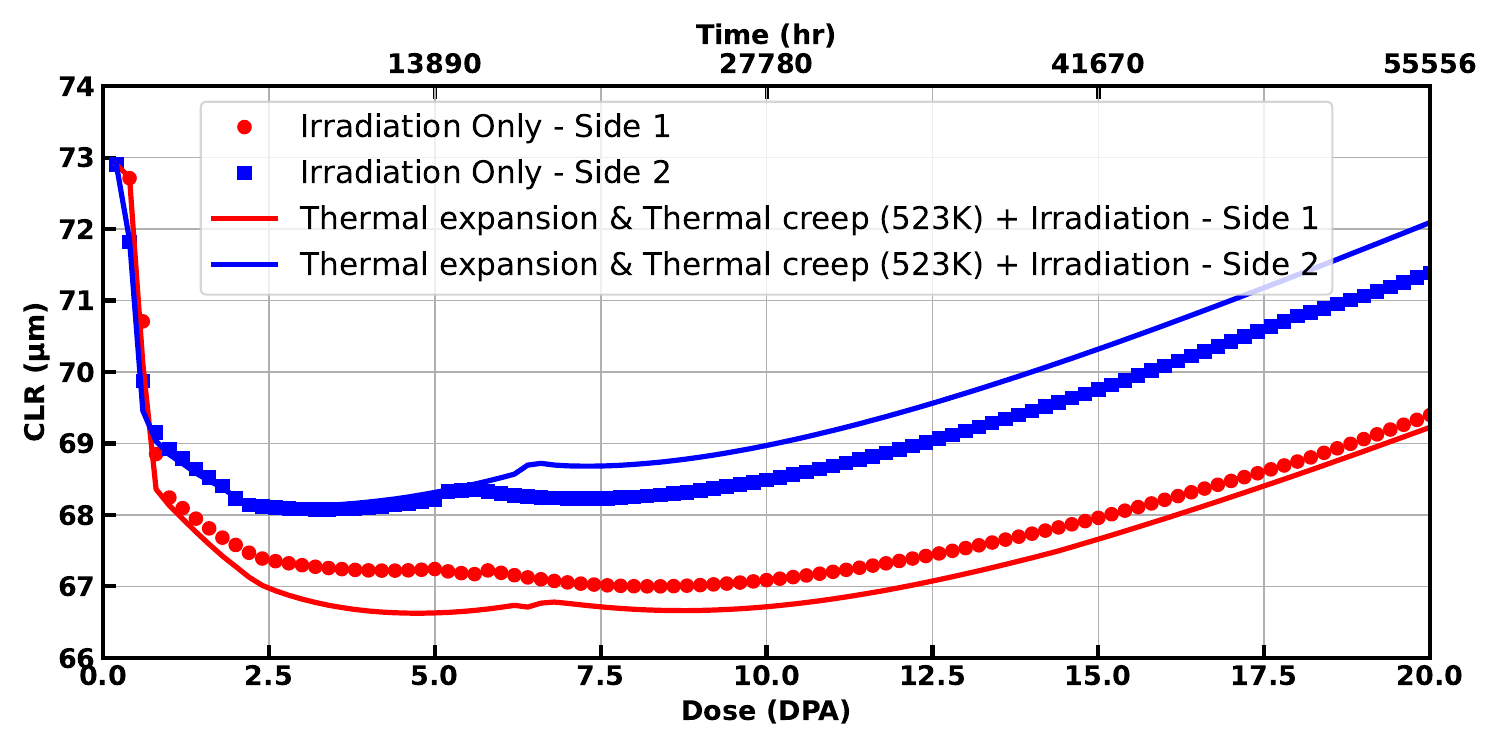}
    \caption{Maximum CLR as a function of dose (DPA), comparing irradiation-only cases (symbols) and combined irradiation + thermal loading (lines), up to 20~dpa (55,560~hours).}
    \label{fig:20dpa}
\end{figure}

\begin{figure}[h!] 
    \centering
    \includegraphics[width=0.5\textwidth]{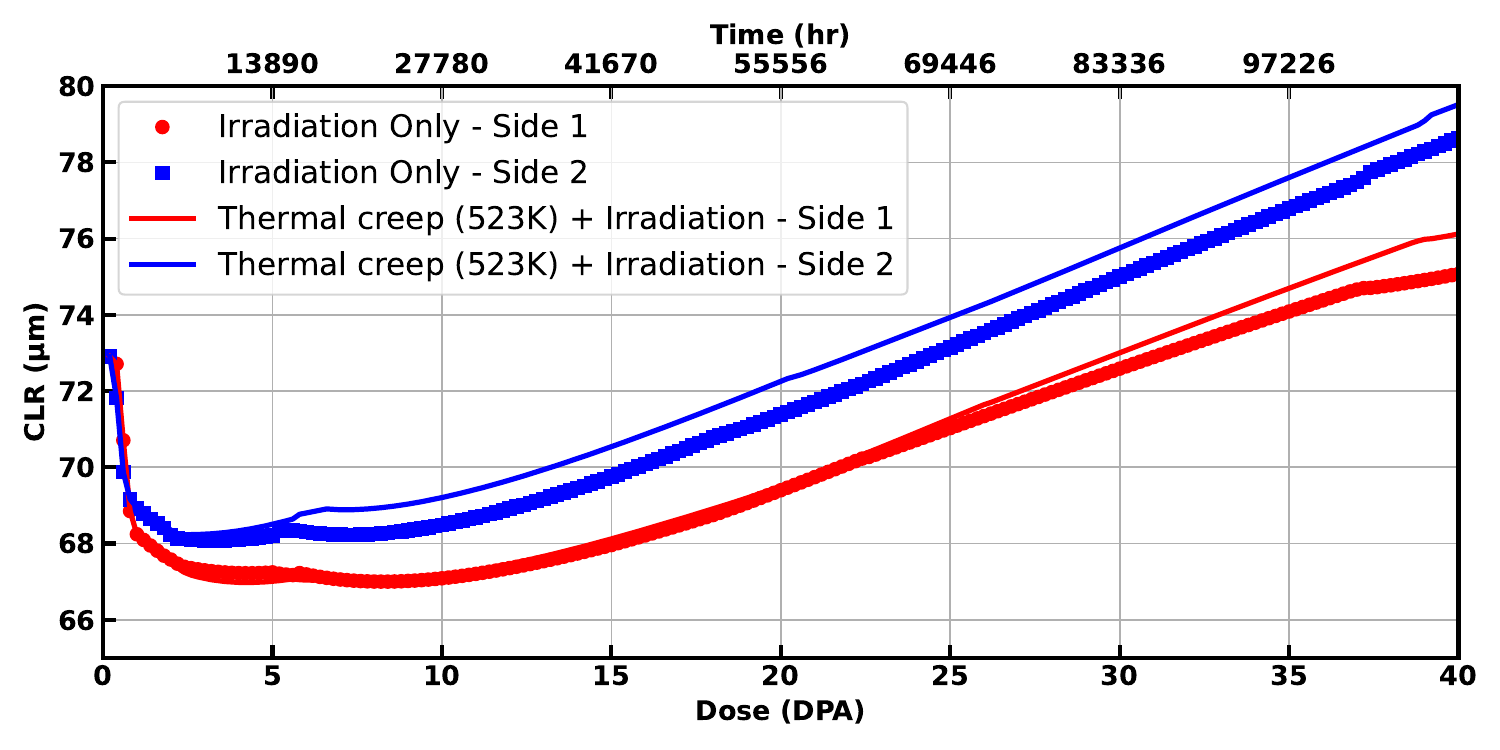}
    \caption{Maximum CLR as a function of dose (DPA), comparing irradiation-only and coupled thermal+irradiation loading, up to 40~dpa (111,120~hours).}
    \label{fig:40dpa}
\end{figure}

Figures~\ref{fig:20dpa} and~\ref{fig:40dpa} further compare the evolution of the CLR at both sides of the spacer grid, considering cases with and without thermal effects. Figure~\ref{fig:20dpa} covers up to 20~dpa (55,556~hours), while Figure~\ref{fig:40dpa} extends the simulation to 40~dpa (111,120~hours).

The curves corresponding to simulations with and without thermal phenomena remain nearly parallel, indicating that thermal expansion and thermal creep introduce only minor deviations of grid and cladding interaction under the conditions studied. These results confirm that irradiation-induced mechanisms, particularly irradiation creep and growth, are the dominant contributors to deformation and separation in the system.

Thermal effects, although limited for the current component geometry and operating conditions, were included to ensure a comprehensive assessment of the system behavior.

Introducing thermal effects into the model increases the computational cost significantly. While the cases considering only irradiation effects required approximately 4 to 6 hours to complete, the fully coupled simulations—accounting for both thermal and irradiation phenomena—took between 14 and 15 hours. All simulations were executed using \texttt{MPI} with 4 parallel processes on an AMD EPYC 7513 CPU running at 2.6~GHz. This increase in computational demand highlights the trade-off between physical fidelity and numerical cost when modeling coupled multiphysics scenarios.

\subsection{Influence of the thermal effects on CLR}
\label{Influence of the thermal effects on CLR}

\begin{figure}[h!] 
    \centering
    \includegraphics[width=0.5\textwidth]{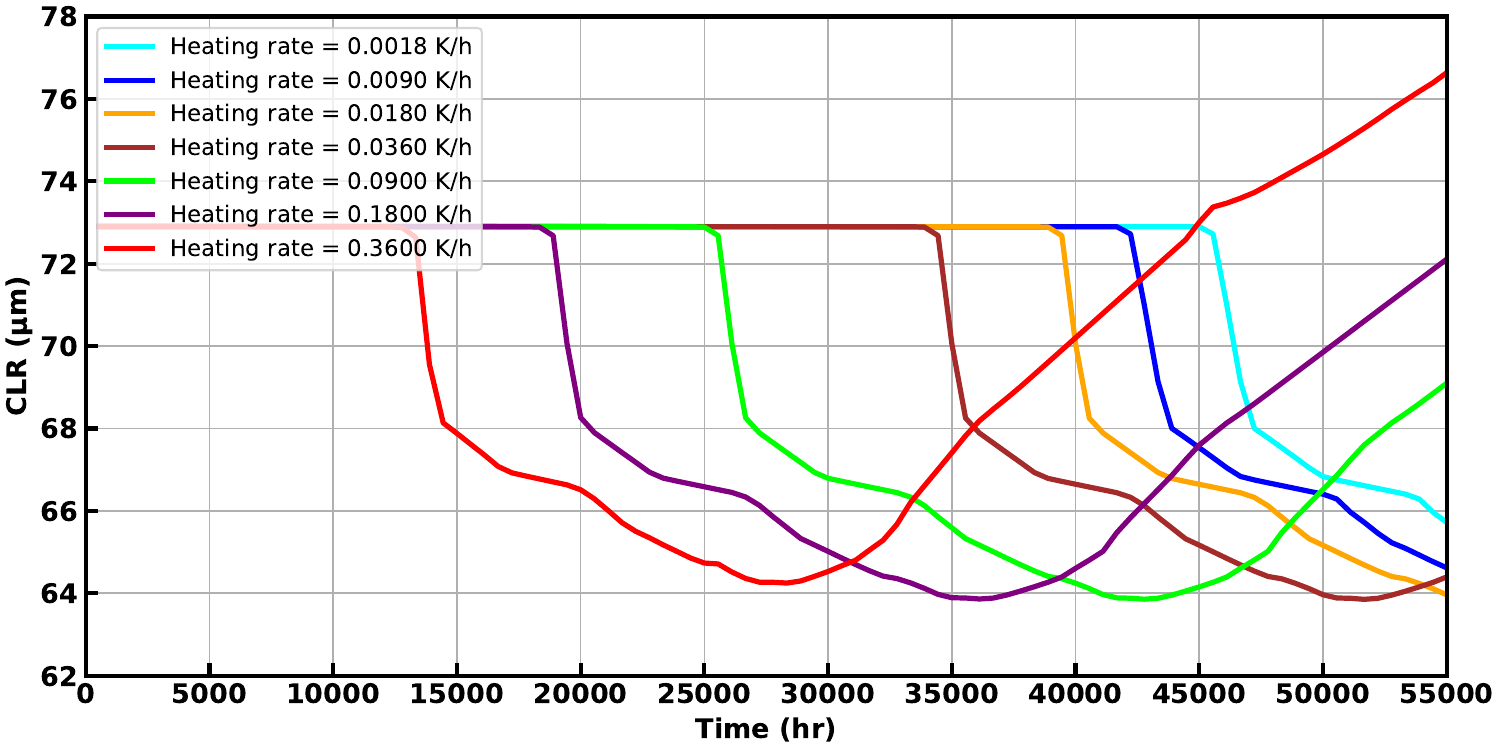}
    \caption{Evolution of CLR under different heating rates considering only thermal expansion. Higher heating rates result in earlier and stronger separation phases.}
    \label{fig:MaxJEU_Test1}
\end{figure}

Two sensitivity analyses are carried out to asses the impact of the heating rate over the assembly mechanics. Figure~\ref{fig:MaxJEU_Test1} presents the evolution of the cladding-to-grid CLR over time considering only thermal expansion for different heating rates. The results indicate that higher ramp rates lead to an earlier onset of instability and contact between the cladding tube and the spacer grid, reaching a minimum CLR of approximately 64~$\mu$m. This is followed by a separation phase where the gap increases. As the heating rate increases, both the timing and steepness of the contact and separation phases become more pronounced.

\begin{figure}[h!] 
    \centering
    \includegraphics[width=0.5\textwidth]{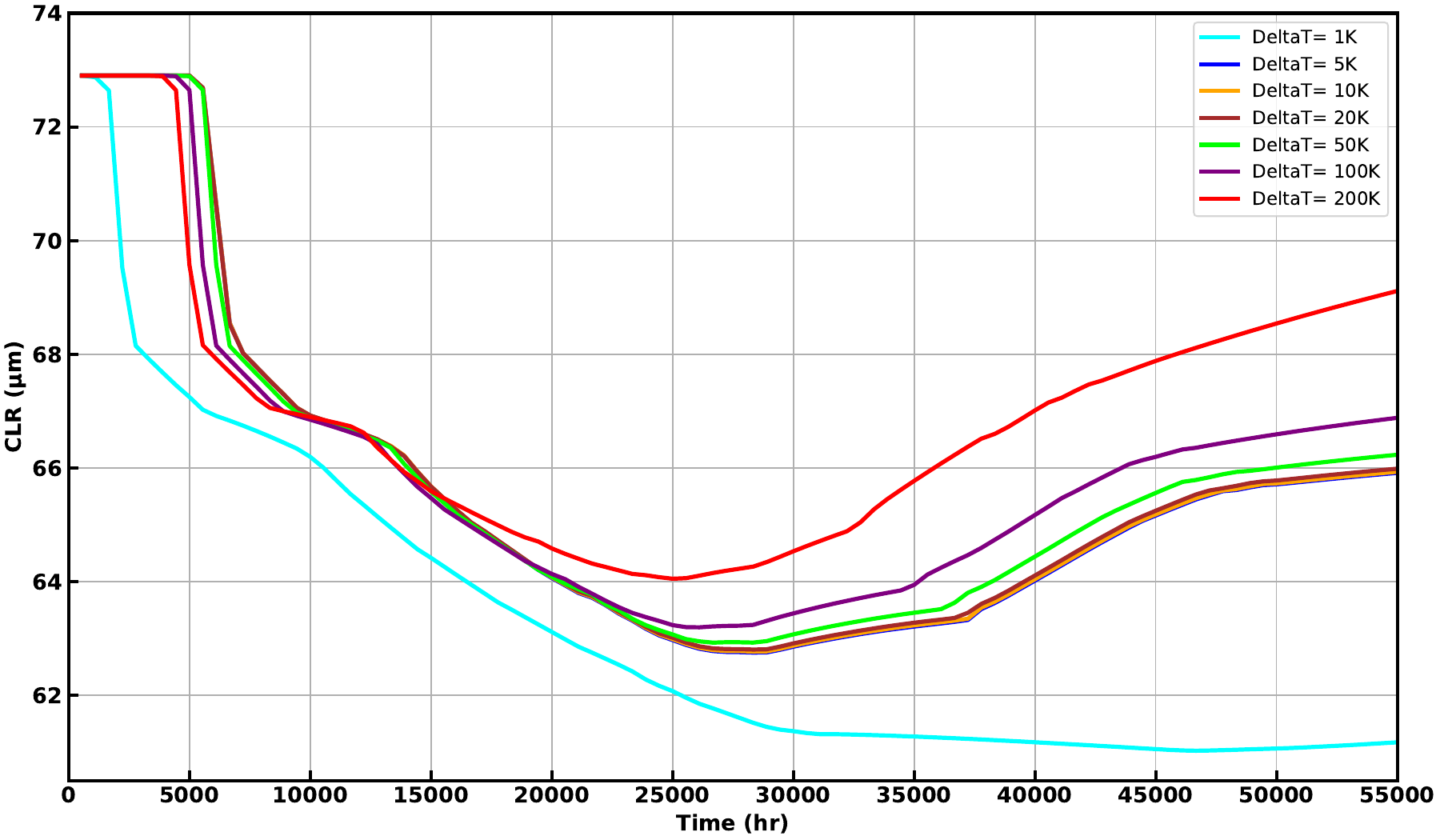}
    \caption{CLR evolution under different heating rates considering only thermal creep. Increased heating rates reduce the overall approach due to enhanced strain rates and the specific crystallographic orientation of the cladding.}
    \label{fig:MaxJEU_Test2}
\end{figure}

Figure~\ref{fig:MaxJEU_Test2} shows the corresponding CLR evolution when only thermal creep is considered under the same heating rate scenarios. In this case, faster heating results in a smaller approach between tube and grid. This behavior is attributed to the higher creep strain rates generated by steeper thermal gradients and the compressive radial stress applied to the tube. Due to the crystallographic texture—particularly the alignment of prismatic planes perpendicular to the radial stress—a faster heating rate favors deformation along directions with lower resolved shear stress, thereby limiting radial displacement and reducing contact with the spacer grid.

\subsubsection{Thermal pre-load effect on irradiated components}
\label{Thermal pre-load effect}

This study compares two simulation scenarios: (1) without thermal pre-loading prior to irradiation, and (2) with thermal pre-loading (\(\Delta T = 523 K\)) applied before irradiation, to evaluate its impact on the CLR evolution between the cladding tube and spacer grid.

This thermal pre-loading scenario is representative of conditions such as hot shutdowns or hot standbys in PWR, where the core remains at high temperature and pressure despite being subcritical. Under these circumstances, fuel components may undergo significant thermal expansion prior to irradiation, potentially influencing early-stage mechanical interaction.

\begin{figure}[h!]
%\vspace{-8.cm} % Adjust this value to fine-tune spacing
\centering
\includegraphics[width=0.5\textwidth]{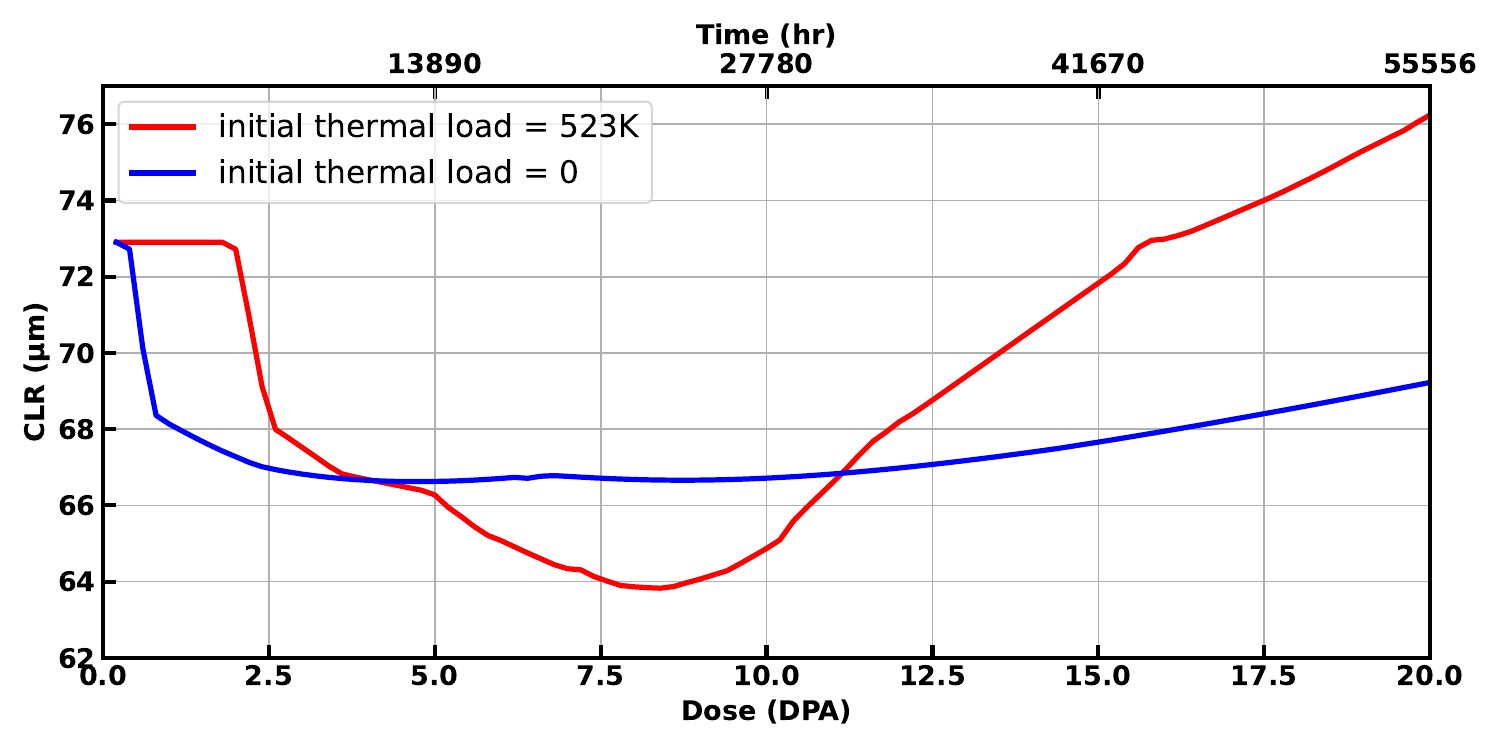}
\caption{Simulated CLR comparison for Side 1: red = with thermal pre-load (\(\Delta T = 523 K\)), blue = without pre-load. Key features: (i) Pre-loaded case shows initial elastic regime (\(<2 dpa\)) followed by creep-driven approach \( \approx 64\,\mu\text{m} \)); (ii) Non-preloaded case exhibits early separation (\(0.5-1.2 dpa\)) then stabilization. Error bands: (\(\pm 5 \mu m\)).}
\label{Fig:ThermalLoad}
\end{figure}

Hence, thermal pre-loading induces initial thermal expansion, maintaining a near-constant CLR during early irradiation stages (up to \(\approx 2 dpa\)), demonstrating predominantly elastic behavior. Beyond this threshold, nonlinearity emerges due to thermal creep softening, resulting in a gradual tube-grid approach (maximum clearance: \( \approx 64\,\mu\text{m} \)). The dimple-tube geometry generates localized stress concentrations (see Fig. \ref{fig:acople}(j-l)), leading to asymmetric deformation with peak CLR (\( \approx 76\,\mu\text{m} \)) at the dimple's opposite edge, as evidenced in Fig.~\ref{Fig:ThermalLoad} (post \(\approx 8 dpa\)).

In contrast, simulations without pre-loading exhibit early pivot-induced separation (initiated at \(\approx 0.5 dpa\), stabilized by \(\approx 1.2 dpa\)), followed by CLR stabilization. While experimental validation remains pending, these results provide some key insights insights for dynamic fuel assembly interaction analysis, grid-to-cladding clearance optimization in design phases, irradiation-enhanced deformation mechanisms.

\subsection{Detailed description of the spatial stress-strain distribution.}

\label{structural analysis}

\begin{figure*}[t!]
	\centering
	
	\textbf{(a)--(f) Global response (cladding + spacer grid)}\par\vspace{0.3em}
	
	% --- Global strain ---
	\begin{subfigure}[b]{0.32\linewidth}
		\includegraphics[width=\linewidth,height=4cm]{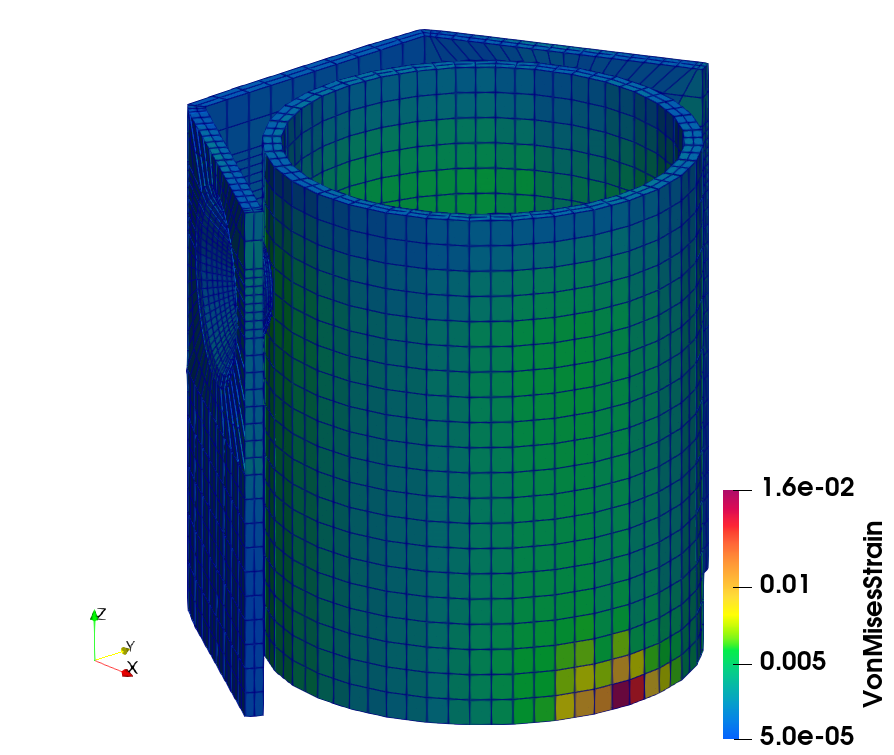}
		\caption{Irradiation-induced strain}
	\end{subfigure}\hfill
	\begin{subfigure}[b]{0.32\linewidth}
		\includegraphics[width=\linewidth,height=4cm]{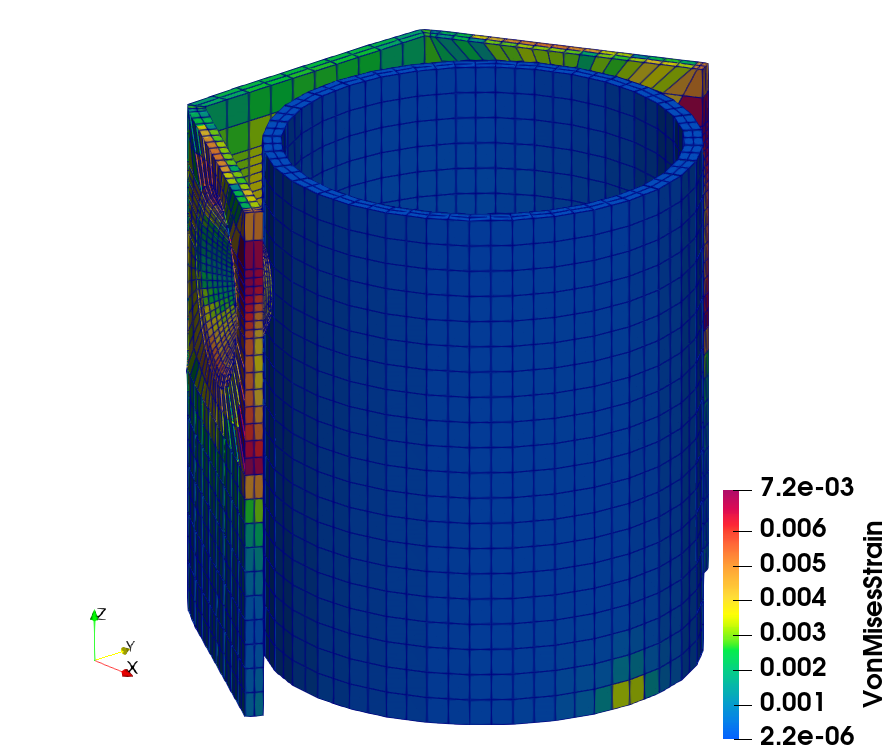}
		\caption{Thermal strain}
	\end{subfigure}\hfill
	\begin{subfigure}[b]{0.32\linewidth}
		\includegraphics[width=\linewidth,height=4cm]{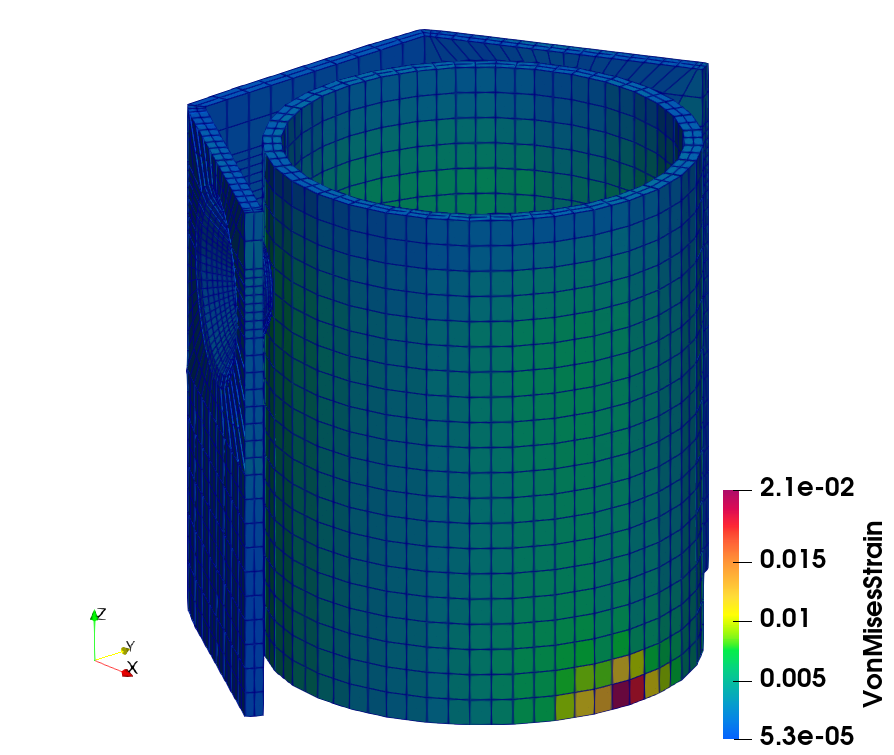}
		\caption{Coupled strain}
	\end{subfigure}
	
	\vspace{0.4em}
	
	% --- Global stress ---
	\begin{subfigure}[b]{0.32\linewidth}
		\includegraphics[width=\linewidth,height=4cm]{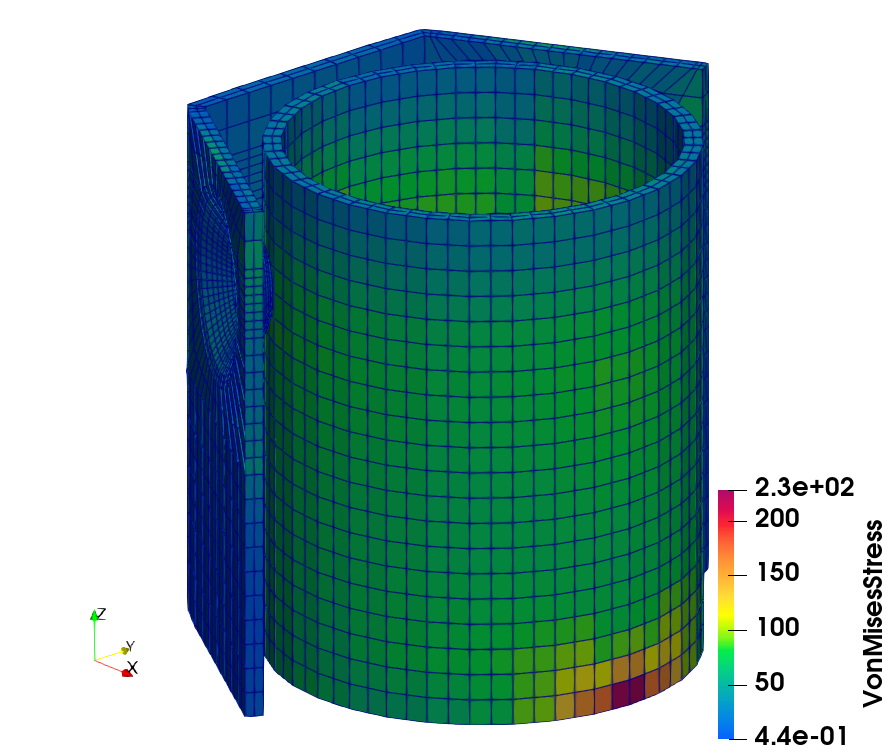}
		\caption{Irradiation stress}
	\end{subfigure}\hfill
	\begin{subfigure}[b]{0.32\linewidth}
		\includegraphics[width=\linewidth,height=4cm]{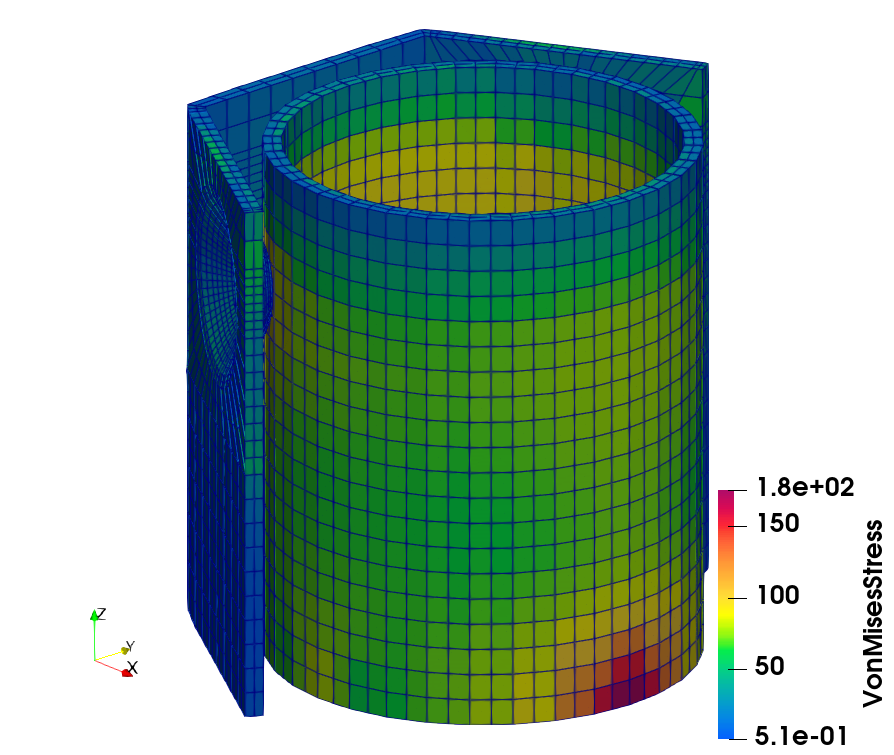}
		\caption{Thermal stress}
	\end{subfigure}\hfill
	\begin{subfigure}[b]{0.32\linewidth}
		\includegraphics[width=\linewidth,height=4cm]{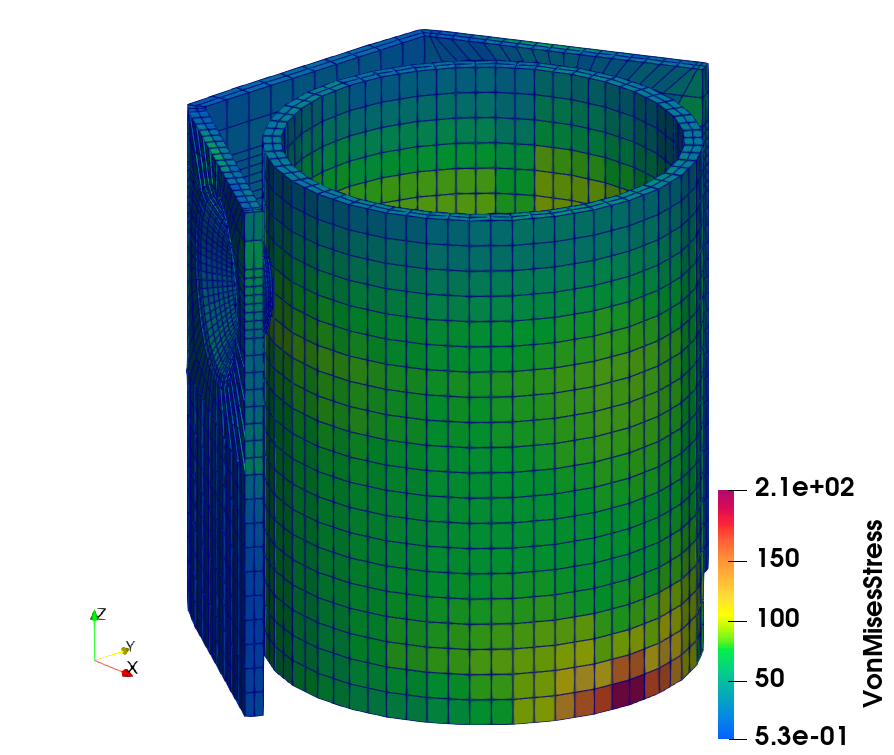}
		\caption{Coupled stress}
	\end{subfigure}
	
	\vspace{0.6em}
	\textbf{(g)--(l) Local response at dimple contact (Side 1)}\par\vspace{0.3em}
	
	% --- Local strain ---
	\begin{subfigure}[b]{0.32\linewidth}
		\includegraphics[width=\linewidth,height=4cm]{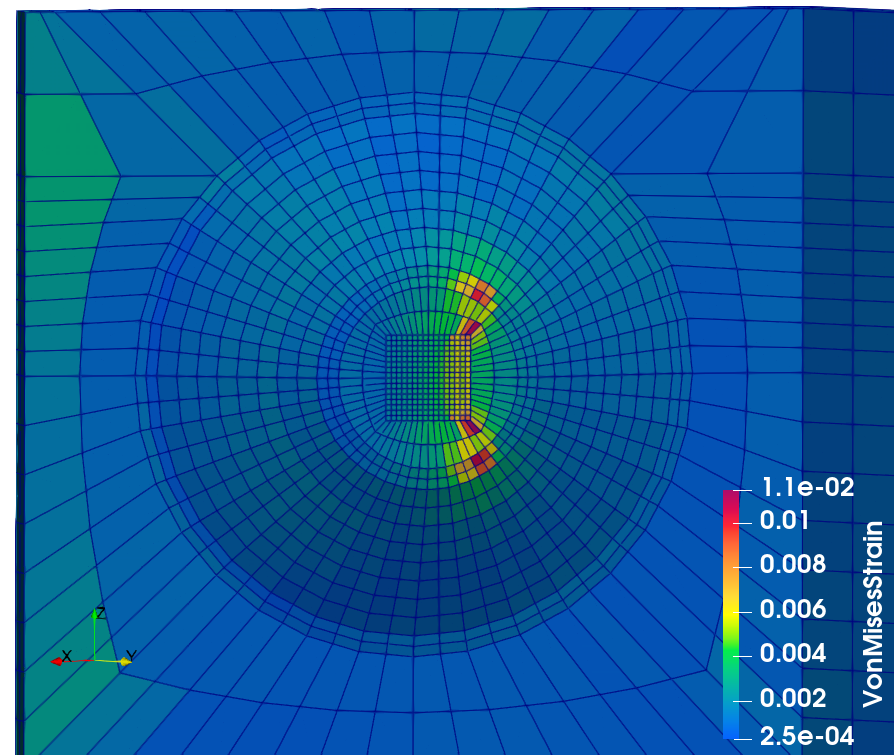}
		\caption{Irradiation-induced strain}
	\end{subfigure}\hfill
	\begin{subfigure}[b]{0.32\linewidth}
		\includegraphics[width=\linewidth,height=4cm]{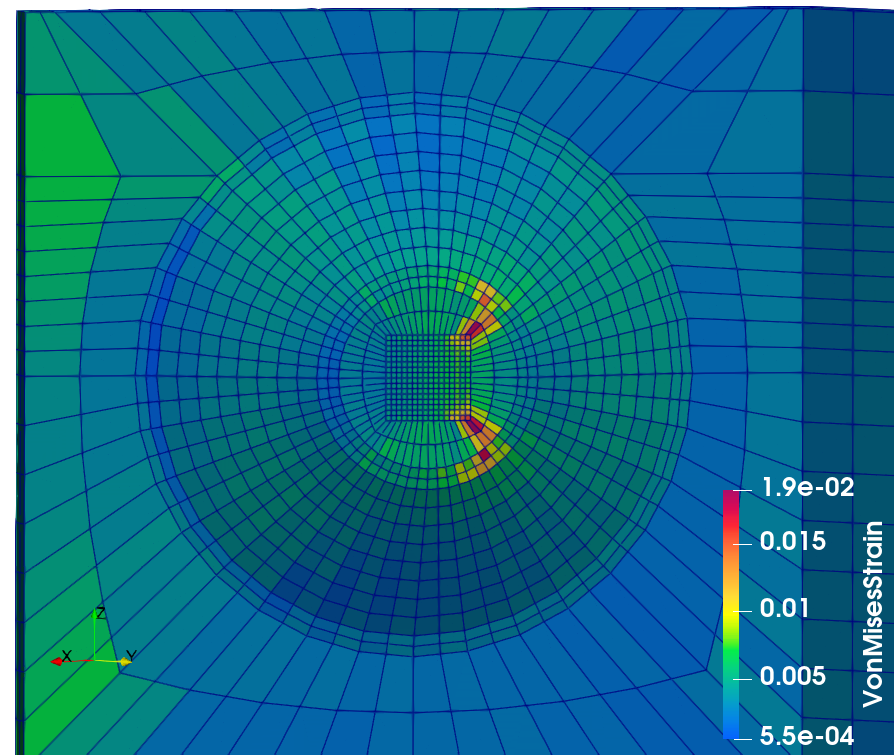}
		\caption{Thermal strain}
	\end{subfigure}\hfill
	\begin{subfigure}[b]{0.32\linewidth}
		\includegraphics[width=\linewidth,height=4cm]{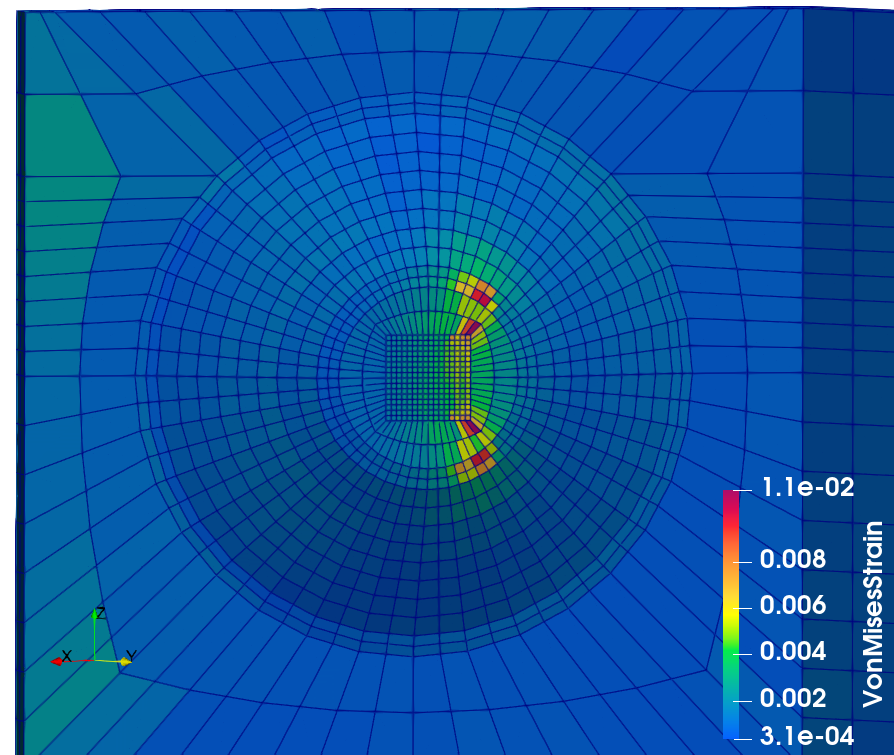}
		\caption{Coupled strain}
	\end{subfigure}
	
	\vspace{0.4em}
	
	% --- Local stress ---
	\begin{subfigure}[b]{0.32\linewidth}
		\includegraphics[width=\linewidth,height=4cm]{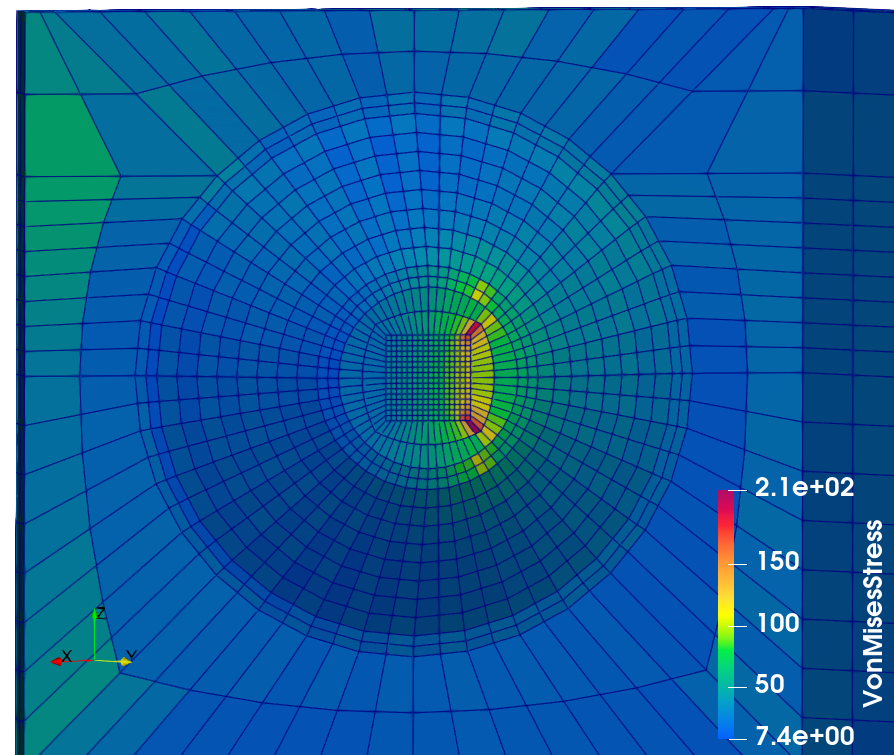}
		\caption{Irradiation stress}
	\end{subfigure}\hfill
	\begin{subfigure}[b]{0.32\linewidth}
		\includegraphics[width=\linewidth,height=4cm]{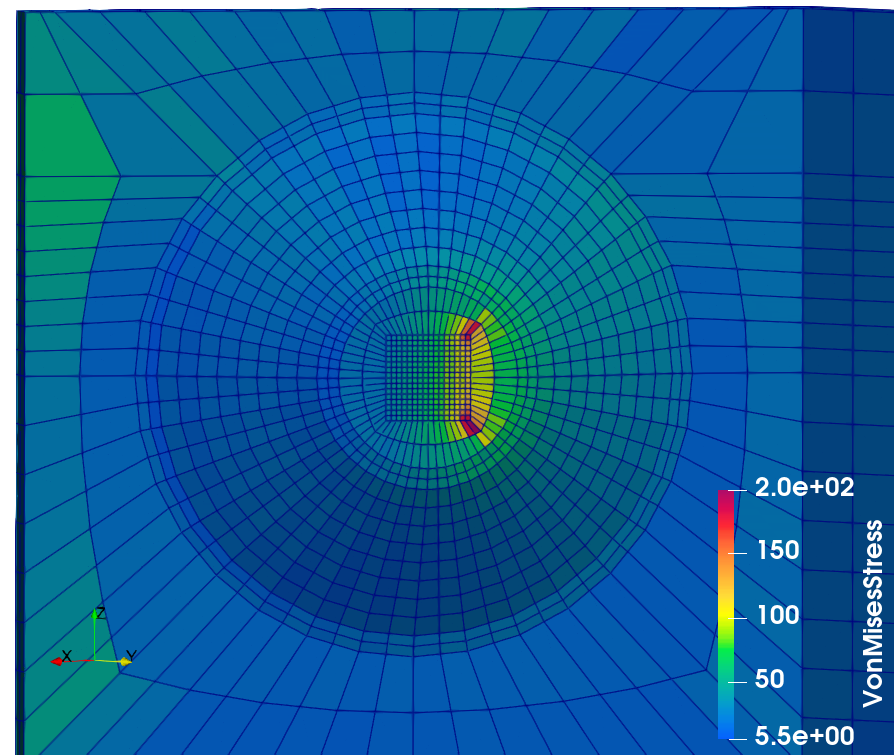}
		\caption{Thermal stress}
	\end{subfigure}\hfill
	\begin{subfigure}[b]{0.32\linewidth}
		\includegraphics[width=\linewidth,height=4cm]{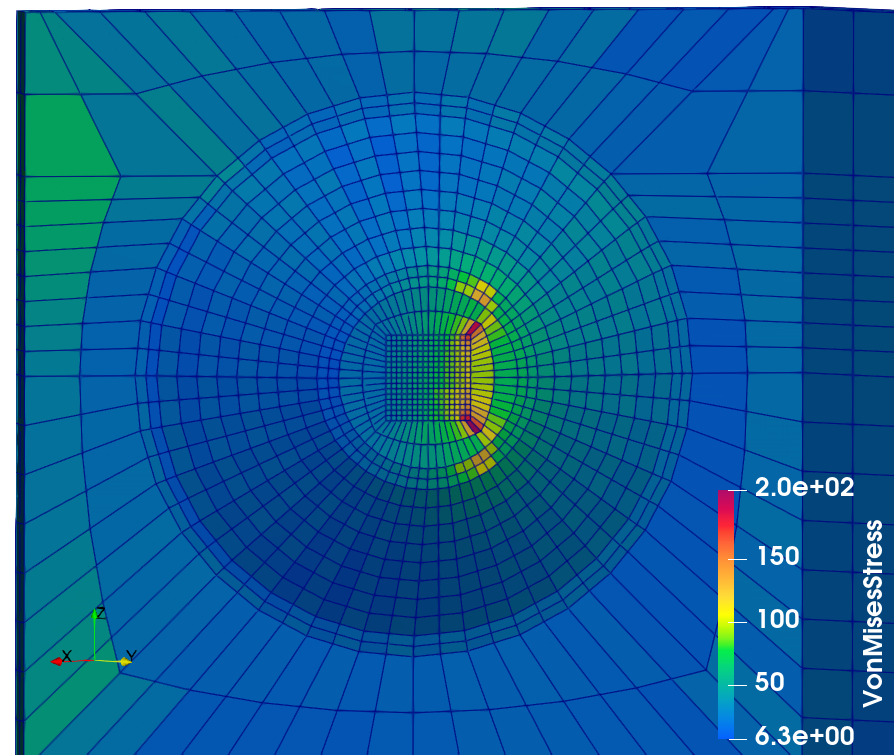}
		\caption{Coupled stress}
	\end{subfigure}
	
	\caption{VM elastic strain and stress distributions under different loading conditions. (a–f): Global response. (g–l): Local response.}
	\label{fig:CombinedAnalysis}
\end{figure*}

A more detailed description of the structural behavior is presented in Figure~\ref{fig:CombinedAnalysis}(a–f), which shows the distribution of von Mises (VM) strains (top row) and VM stresses (bottom row) under combined thermal and irradiation conditions. The analysis is divided into three cases: the first column corresponds to the scenario with irradiation effects only (irradiation creep and growth), the second column includes only thermal effects (thermal creep and thermal expansion with heating rate $=0.0018$~K/h), and the third column represents the superposition of all phenomena.

When only irradiation mechanisms are considered, a localized plastic deformation of approximately \(1.6 \times 10^{-2}\) develops specifically in the region where the spring load is applied on the cadding tube, accompanied by VM stress concentrations reaching nearly 230~MPa. In contrast, when only thermal effects (thermal expansion and creep) are included, the deformation is significantly lower, reaching about \(7.2 \times 10^{-3}\), with a corresponding VM stress of approximately 180~MPa. These results confirm the relatively minor contribution of thermal effects under the given loading conditions.

When all mechanisms act simultaneously—irradiation growth and creep, thermal expansion, and thermal creep—the resulting deformation in the loaded region increases to approximately \(2.1 \times 10^{-2}\), representing a 31\% increase in strain compared to the irradiation-only case. Meanwhile, the VM stress slightly decreases to 210~MPa in the spring load region. This suggests a localized stress relaxation effect, likely driven by the interplay between irradiation-induced strain and thermally activated material softening, which facilitates stress redistribution despite continued strain accumulation.

\vspace{-1pt} % Compensación después de la figura
Figure~\ref{fig:CombinedAnalysis}(g–l) presents the local VM strain (top row) and stress (bottom row) distributions at the dimple contact region on Side~1, under different loading conditions. These local results complement the global response shown in (a–f). The first column shows the outcome with only irradiation effects (irradiation creep and growth), the second with only thermal effects (thermal creep and thermal expansion), and the third with both irradiation and thermal mechanisms combined.

When only irradiation effects are considered (first column), a localized strain of approximately $1.1 \times 10^{-2}$ is observed at the contact points between the cladding tube and the grid dimples. In the same regions, VM stresses reach around 210~MPa. This behavior reflects the accumulation of irradiation-induced strain in the high-contact stress zones.

In contrast, when only thermal effects are considered (second column), the maximum localized strain is greater, reaching approximately \(1.9 \times 10^{-2}\), while the VM stress slightly decreases to around 200~MPa.

When both thermal and irradiation effects are taken into account simultaneously (third column), the deformation field is dominated by the irradiation-induced strain, while thermal effects contribute mainly to the relaxation of stresses in the contact region.

%%%%%%%%%%%%%%%%%%%%%%%%%%%%%%%%%%%%%%%%%%%%%%%%%%5
%%%%%%%%%%%%%%%%%%%%%%%%%%%%%%%%%%%%%%%%%%%%%%%%%%5

\subsubsection{Influence of elastic coefficients on stress and strain fields}
\
\label{hot elastic constants}
The anisotropic elastic constants of zirconium as a function of temperature were determined by \cite{fisher1961adiabatic,fisher1964single} in a range spanning from 4~K to approximately 1100~K. Additionally, \cite{tremblay1973elastic} studied the elastic constants of zirconium in an alloy with high oxygen content. It is well known that the presence of alloying elements modifies the lattice parameters and, consequently, affects the elastic constants.

In this work, the values reported by \cite{douglass1971metallurgy} were used for the simulation of thermal expansion. At a temperature of 576~K, the elastic constants employed were:
\begin{equation*}
	\begin{aligned}
		C_{11} &= 130.1 \ \text{GPa}, \quad
		C_{33} = 155.9 \ \text{GPa}, \\
		C_{44} &= 28.2 \ \text{GPa}, \quad
		C_{12} = 78.6 \ \text{GPa}, \quad
		C_{13} = 65.7 \ \text{GPa}.
	\end{aligned}
\end{equation*}

\begin{figure*}[t!]
	\centering
	
	\textbf{(a)--(b) Elastic strain fields \qquad (c)--(d) Elastic stress fields}\par
	\vspace{0.5em}
	
	% Row 1 - Strains
	\begin{subfigure}[t]{0.45\linewidth}
		\centering
		\includegraphics[height=4.2cm]{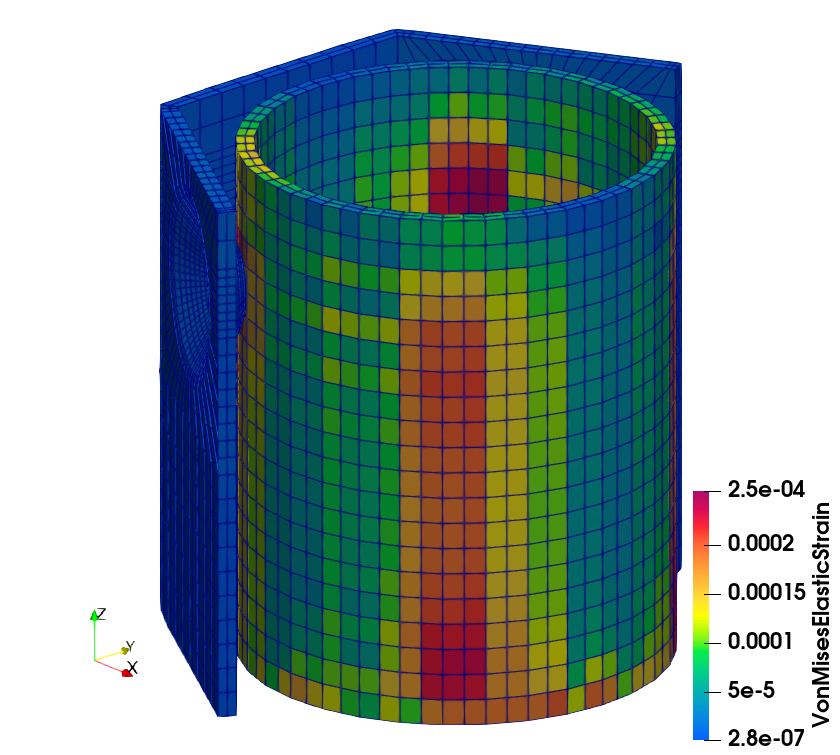}
		\caption{VM elastic strain (cold-state coefficients)}
		\label{subfig:strain_cold}
	\end{subfigure}
	\hspace{0.05\linewidth}
	\begin{subfigure}[t]{0.45\linewidth}
		\centering
		\includegraphics[height=4.2cm]{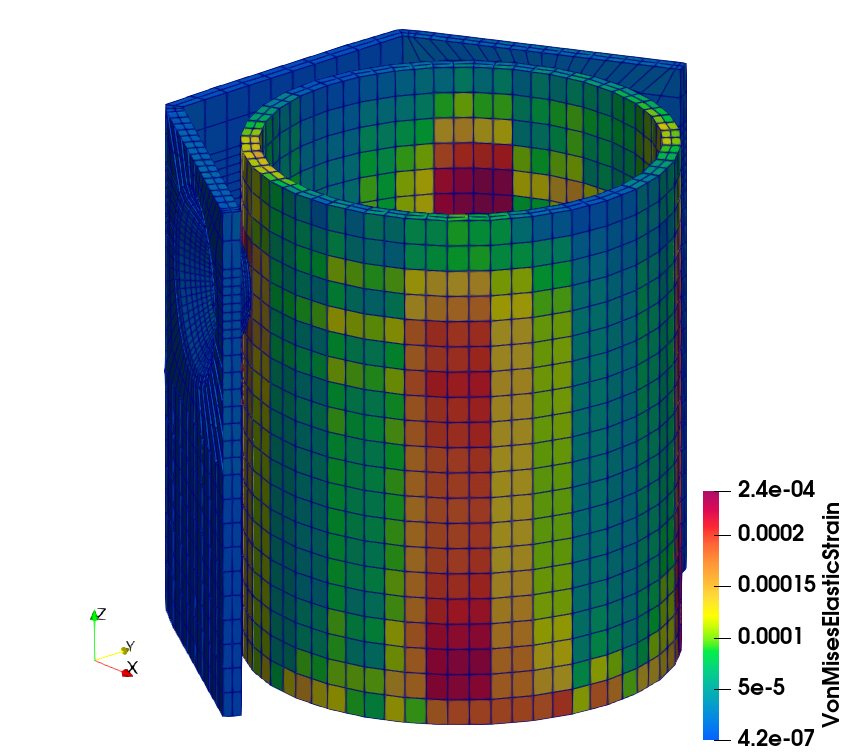}
		\caption{VM elastic strain (576~K coefficients)}
		\label{subfig:strain_hot}
	\end{subfigure}
	
	\vspace{0.5em}
	
	% Row 2 - Stresses
	\begin{subfigure}[t]{0.45\linewidth}
		\centering
		\includegraphics[height=4.2cm]{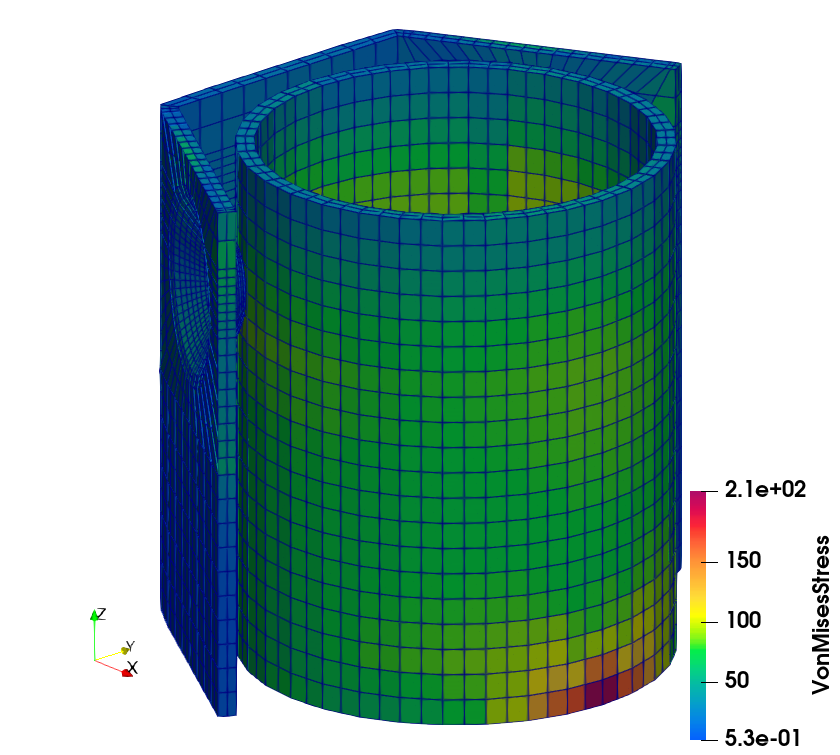}
		\caption{VM stress (cold-state coefficients)}
		\label{subfig:stress_cold}
	\end{subfigure}
	\hspace{0.05\linewidth}
	\begin{subfigure}[t]{0.45\linewidth}
		\centering
		\includegraphics[height=4.2cm]{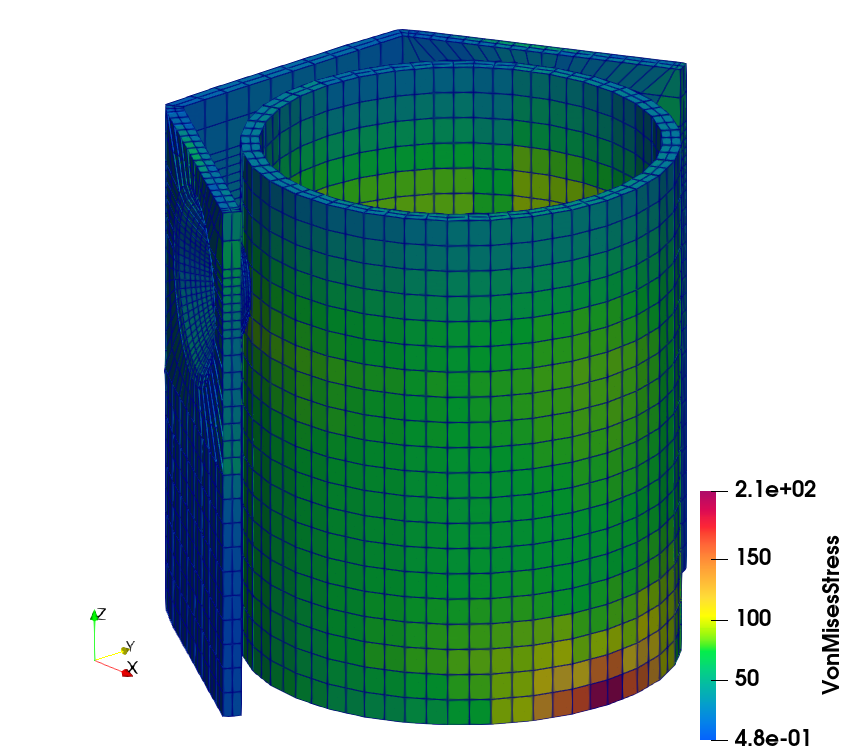}
		\caption{VM stress (576~K coefficients)}
		\label{subfig:stress_hot}
	\end{subfigure}
	
	\captionsetup{justification=justified,font=small}
	\caption{
		Global distributions of VM elastic strain and stress in the full assembly (cladding tube + spacer grid) under coupled irradiation and thermal loading.  
		(a–b): Strain fields using elastic coefficients at room temperature and 576~K, respectively.  
		(c–d): Corresponding stress fields. Results highlight the influence of elastic moduli selection on local deformation and stress distribution.
	}
	\label{fig:Elastic Behavior}
\end{figure*}

Figures~\ref{fig:Elastic Behavior} present the distributions of VM elastic strains and stresses for two cases: (1) using room temperature elastic constants (see Section~\ref{model due to Thermal Expansion}), and (2) using elastic constants at 576~K (see Section~\ref{hot elastic constants}). A component-level comparison shows that the equivalent elastic strain reaches approximately $2.5 \times 10^{-4}$ in the first case and $2.4 \times 10^{-4}$ in the second. The equivalent VM stress remains nearly identical in both simulations, with a maximum of approximately 210~MPa.

\begin{figure}[htbp]
	\centering
	\includegraphics[width=0.5\textwidth]{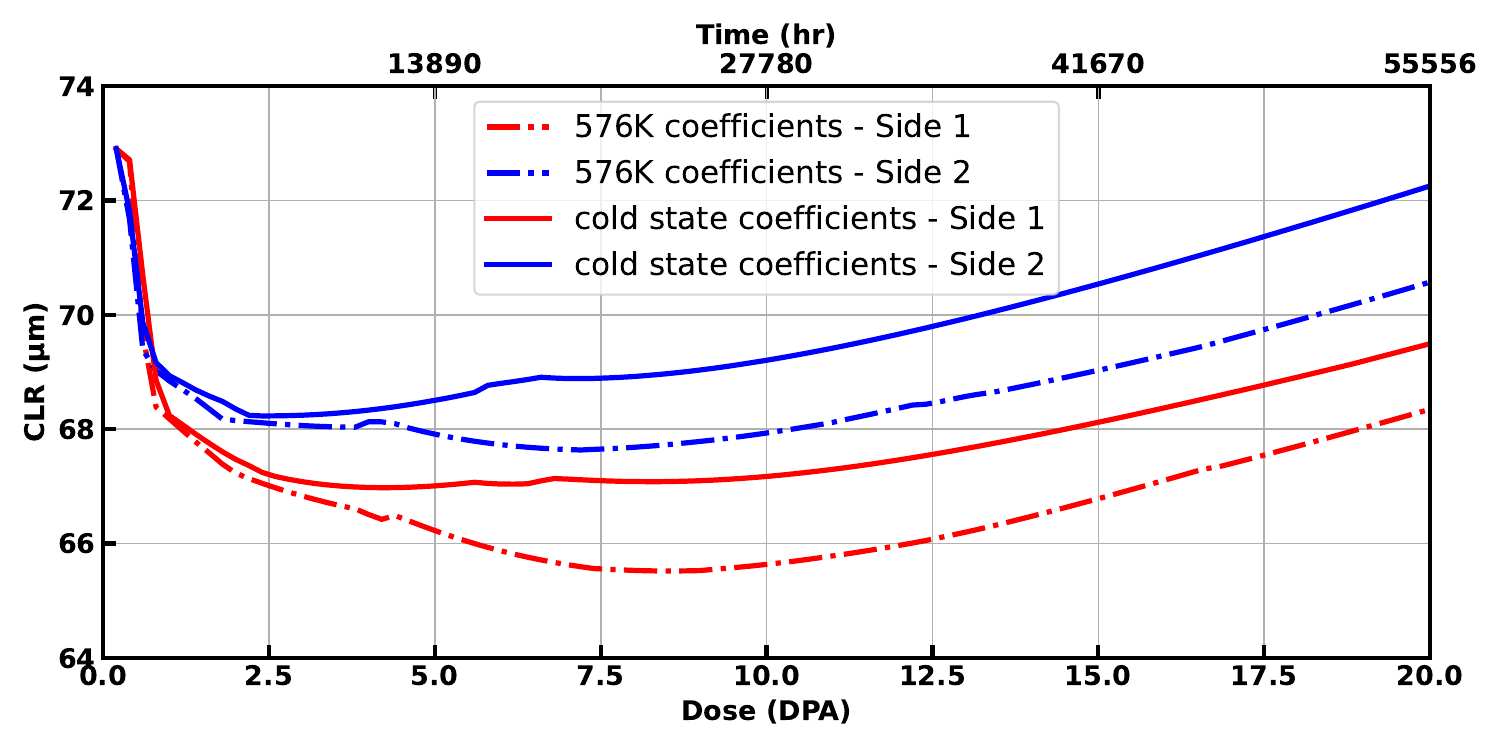}
	\caption{Influence of elastic coefficients on CLR between the cladding tube and spacer grid.}
	\label{fig:CoefColdvsCoefHot}
\end{figure}

In particular, the comparison suggests that the thermal efffects on the the elastic constants contribute to stress relaxation and increase deformation on the early stages (see Fig.~\ref{fig:CombinedAnalysis}d--f). This can be explained by means of the reduction of the elastic coefficients corresponding to 576 K, that promote greater localized deformation compared to those of the cold-worked state. This indicates that temperature impacts the material response mainly through micromechanical phenomena, rather than through the temperature dependence of the elastic constants alone. This effect is further illustrated in Figure~\ref{fig:CoefColdvsCoefHot}, which compares the CLR between the cladding tube and spacer grid for both sets of elastic coefficients. While the stress and strain fields show negligible differences, the use of coefficients at 576~K produces a relative approach of approximately 2~µm, suggesting a non-negligible influence of elastic stiffness on contact evolution.

%%%%%%%%%%%%%%%%%%%%%%%%%%%%%%%%%%%%%%%%%%%%%%%%%%5
%%%%%%%%%%%%%%%%%%%%%%%%%%%%%%%%%%%%%%%%%%%%%%%%%%5

%Additional space adjustment if needed
% Primera figura (vista general)

\section{Conclusion}
\label{sec:conclusion}

This work incorporates thermally activated mechanisms such as creep and expansion within the CAFEM-VPSC interface developed in \citep{aguzzi2025toolbox} to simulate irradiated component assemblies under typical PWR conditions. The numerical results support the following conclusions:

\begin{enumerate}
	
	\item \textbf{Thermal creep is stress-sensitive and leads to higher strain under steady-state conditions:}  
	The analysis confirms that, for a given stress level, total strain increases consistently when thermal effects are included. At 523~K under an axial load of 200~MPa, the axial strain rises from 0.018 (only irradiation effects at 20~dpa) to 0.027 when thermal creep and expansion are also considered—representing a 50\% increase. This highlights the stress-dependent nature of thermal creep and its relevance for modeling cladding deformation under standard reactor operating conditions.
	
	\item \textbf{Thermal effect impact to CLR evolution:}  
	For this specific spacer grid–cladding configuration and operation conditions, thermally induced deformations (creep and expansion) exhibit a very small influence on contact dynamics, with CLR variations remaining below 1\% compared to irradiation-dominated scenarios. However, thermal preloading (e.g., $\Delta$T = 523 K) alters the initial gap stability, accelerating tube-to-grid contact recovery. This early contact onset could enhance mechanical stability by mitigating flow-induced vibration risks during initial reactor operation. This phenomenon would be relevant within a modal analysis on these components.
	
	\item \textbf{Stress relaxation under full coupling:}  
	The structural analysis (Section~\ref{structural analysis}) showed that the inclusion of thermal effects results in higher von Mises strains but lower stress levels compared to the irradiation-only case. This behavior is attributed to material softening and stress relaxation driven by thermal activation, particularly due to the predominant orientation of prismatic slip systems aligned with the radial load.

	\item \textbf{Strain–stress fields driven by micromechanics over elasticity:}  
	The simulated strain and stress fields were more sensitive to micromechanical behavior (slip activity and defect accumulation) than to the choice of elastic constants (Section~\ref{hot elastic constants}). This is partly influenced by the boundary condition strategy used: the internal nodes on the free end of the cladding were tied to a fictitious remote node, which facilitates displacement toward the grid under spring loading. Future work should focus on calibrating the stiffness of this constraint to improve physical representativeness.
\end{enumerate}

This crystal plasticity-based methodology provides significant insight into the stress–strain response and mechanical interaction of fuel components. Even though in the current application the clearance could be properly estimated by iraddiation models, the thermal effects impact significantly in the local integrity of the components. Hence, the full model serves as a \textit{computational tool} capable of extrapolating microscale results to phenomenological models with lower computational demand. This practice opens possibilities for simulating more complex reactor internal geometries in industrial applications, achieving a realistic approximation of material behavior with significantly reduced time and cost.

\section*{CRediT authorship contribution statement}
F. E. Aguzzi: Methodology. Software, Formal analysis, Writing.  S. M. Rabazzi: Software, Formal analysis. M. S. Armoa: Methodology, Writing. C. I Pairetti: Supervision, Formal analysis, Writing - review \& editing. A. E. Albanesi: Supervision, Formal analysis, Writing - review \& editing.

\section*{Declaration of competing interest}

The authors declare that they have no known competing financial interests or personal relationships that could have appeared to influence the work reported in this paper.

\section*{Acknowledgments}
The authors gratefully acknowledge the financial support from CONICET (Argentine Council for Scientific and Technical Research) and the National Agency of Scientific and Technological Promotion of Argentina (ANPCYT) for the Grant PICT-2020-SERIEA-03475. The authors also acknowledge computer time provided by CCT-Rosario Computational Center, member of the High Performance Computing National System (SNCAD, ME-Argentina).

\FloatBarrier
\section*{Appendix A: Material parameters} 
\renewcommand{\theequation}{A\arabic{equation}}
\renewcommand{\thesection}{A}
\refstepcounter{section} 
\label{apéndiceA}
\FloatBarrier

The material parameters presented in Table \ref{tab:zircaloy_params} were taken from \cite{patra2017crystal} and \cite{patra2017finite}.

\begin{table*}[h]
	\small
	\centering
	\caption{Model parameters for Zircaloy-2, from \cite{patra2017crystal} and \cite{patra2017finite}.}
	\renewcommand{\arraystretch}{1.2}
	\begin{tabular}{lp{6cm}p{8cm}}  % Ajusto el ancho de la segunda y tercera columna
		\toprule
		\textbf{Parameter} & \textbf{Value} & \textbf{Meaning} \\
		\midrule
		$C_{11}, C_{22}, C_{33}$ & 143.5, 143.5, 164.9 & \multirow{2}{=}{Elastic constants in GPa (assumed to be the same as that for pure Zr). Values taken from \cite{simmons1965single} and \cite{kocks2000texture}.} \\
		$C_{12}, C_{13}, C_{23}$ & 72.5, 65.4, 65.4 & \\
		$C_{44}, C_{55}, C_{66}$ & 32.1, 32.1, 35.5 & \\
		\midrule
		$f_r$ & 0.97 & Fraction of point defects that recombine during the cascade \\
		$f_{ic}$ & 0.13 & Fraction of interstitials that form clusters \\
		$B$ & $5.0 \times 10^{-5}$ MPa$\cdot$dpa$^{-1}$ & Crystallographic irradiation creep compliance \\
		$p_{ref}$ & $2.26 \times 10^{14}$ m$^{-2}$ & Weighting factor for line dislocation density in irradiation creep model \\
		$b^i, \ j = \alpha_1, \alpha_2, \alpha_3$ & $3.0 \times 10^{-10}$ m & Magnitude of Burgers vector along the prismatic directions \\
		$b^i, \ j = c$ & $5.0 \times 10^{-10}$ m & Magnitude of Burgers vector along the basal direction \\
		\bottomrule
	\end{tabular}
	\label{tab:zircaloy_params}
\end{table*}

\FloatBarrier
\section*{Appendix B: Validation benchmark} 
\renewcommand{\theequation}{A\arabic{equation}}
\renewcommand{\thesection}{B}
\refstepcounter{section} 
\label{apéndiceB}
\FloatBarrier

\begin{figure}[htbp] 
	\centering
	\includegraphics[width=0.45\textwidth]{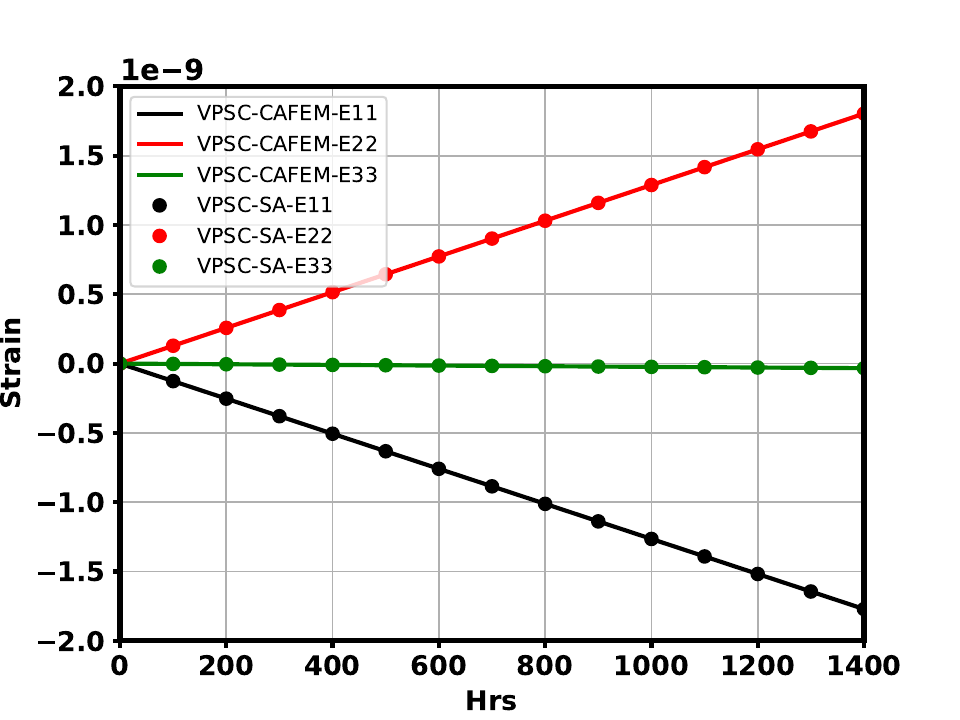}
	\caption{Thermal strain at 523~K for single crystal with 10~MPa applied along direction 22. Symbols represent standalone VPSC results; solid lines correspond to Code Aster–VPSC predictions.}
	\label{fig:ThermalCreepStrain}
\end{figure}

A benchmark case was used to verify the implementation of the thermal creep phenomenon in the Code\_Aster–VPSC coupling. This involved a code-to-code comparison against results obtained with the stand-alone VPSC code. The conditions of the reference case are summarized in Table~\ref{tab:thermal_creep_conditions}. The simulation was performed using a single three-dimensional eight-node finite element under constant stress. The results of the comparison between Code\_Aster–VPSC and stand-alone VPSC are shown in Figure~\ref{fig:ThermalCreepStrain}.

\begin{table}[htbp]
	\centering
	\caption{Simulation conditions for thermal creep tests.}
	\small
	\renewcommand{\arraystretch}{1.2}
	\begin{tabularx}{\columnwidth}{l l l X}
		\toprule
		\textbf{Description} & \textbf{Temp. (K)} & \textbf{Stress (MPa)} & \textbf{Texture} \\
		\midrule
		Thermal Creep & 523 & 10 (22) & (0,0,0)\textsuperscript{*} \\
		\bottomrule
	\end{tabularx}
	\vspace{0.5em}
	\raggedright
	\textsuperscript{*}\small Euler angles: rotation around Z, then X, then Z axis.
	\label{tab:thermal_creep_conditions}
\end{table}

% Figura

%\bibliography{referencias.bib}
\bibliographystyle{cas-model2-names}
\FloatBarrier

\end{document}